\documentclass[twocolumn]{aastex701}

\usepackage{chemformula}
\usepackage{multirow}
\usepackage{listings}
\usepackage{longtable}
\usepackage{placeins}
\usepackage{xcolor}
\newcommand{\rev}[1]{#1}

\usepackage{xspace}
\newcommand\chili{\textsc{CHILI}\xspace}

\begin{document}
\title{Coupled atmospHere Interior modeL Intercomparison (CHILI)\\ Protocol Version 1.0: A CUISINES Intercomparison Project of Magma Ocean Models}

\author[0000-0002-3286-7683]{Tim Lichtenberg}
\email[show]{CHILI\_admin@rockyworlds.org}
\affiliation{Kapteyn Astronomical Institute, University of Groningen, P.O. Box 800, 9700 AV Groningen, The Netherlands}

\author[0000-0003-2915-5025]{Laura Schaefer}
\affiliation{Department of Earth \& Planetary Sciences, Stanford University, Stanford, CA 94305}
\email{lkschaef@stanford.edu}

\author[0000-0001-6878-4866]{Joshua Krissansen-Totton}
\affiliation{Department of Earth and Space Sciences / Astrobiology Program, 
University of Washington, 
Seattle, WA 98105, USA}
\email{joshkt@uw.edu}

\author[0000-0002-0747-8862]{Yamila Miguel}
\affiliation{Leiden Observatory, Einsteinweg 55, 2333 CC Leiden, The Netherlands}
\affiliation{SRON Netherlands Institute for Space Research, Leiden, The Netherlands}
\email{ymiguel@strw.leidenuniv.nl}

\author[0000-0001-8832-5288]{Denis E. Sergeev}
\affiliation{School of Physics, University of Bristol, HH Wills Physics Laboratory, Tyndall Avenue, Bristol BS8 1TL, UK}
\email{denis.sergeev@bristol.ac.uk}

\author[0000-0001-9284-0143]{Philipp Baumeister}
\affiliation{Department of Earth Sciences, Freie Universität Berlin, Malteserstrasse 74-100, 12249 Berlin, Germany}
\email{philipp.baumeister@fu-berlin.de}

\author[0009-0003-5970-570X]{Jessica Cmiel}
\affiliation{Harvard Paulson School of Engineering and Applied Sciences, 29 Oxford Street, Cambridge, MA 02138, USA}
\email{jessicacmiel@g.harvard.edu}

\author[0009-0002-9902-731X]{Leoni J. Janssen}
\affiliation{Leiden Observatory, Einsteinweg 55, 2333 CC Leiden, The Netherlands}
\email{ljanssen@strw.leidenuniv.nl}

\author[0000-0001-6171-200X]{T. Giang Nguyen}
\affiliation{Department of Physics, McGill University, 3600 rue University, Montr\'eal, QC H3A 2T8, Canada}
\email{giang.nguyen@mcgill.ca}

\author[0000-0001-6878-4866]{Yoshinori Miyazaki}
\affiliation{Department of Earth and Planetary Sciences, Rutgers University, 
Piscataway, NJ 08854, USA}
\email{yoshi.miyazaki@rutgers.edu}

\author[0000-0002-8368-4641]{Harrison Nicholls}
\affiliation{Institute of Astronomy, University of Cambridge, Madingley Road, Cambridge CB3 0HA, United Kingdom}
\affiliation{Atmospheric Oceanic and Planetary Physics, University of Oxford, Parks Road, Oxford OX1 3PU, United Kingdom}
\email{harrison.nicholls@ast.cam.ox.ac.uk}

\author[0009-0002-0821-062X]{Alexandra Papesh}
\affiliation{Department of Earth and Space Sciences / Astrobiology Program, University of Washington, Seattle, WA 98105, USA}
\email{agp99@uw.edu}

\author[0009-0005-5234-6673]{Hugo Pelissard}
\affiliation{Laboratoire d’astrophysique de Bordeaux, Univ. Bordeaux, CNRS, B18N, allée Geoffroy Saint-Hilaire, 33615 Pessac, France}
\email{hugo.pelissard@u-bordeaux.fr}

\author[0009-0009-6098-296X]{Bo Peng}
\affiliation{Stanford University, Department of Earth and Planetary Sciences, 450 Jane Stanford Way, Stanford, 94305, CA, USA}
\email{bpengeps@stanford.edu}

\author[0000-0002-9032-8530]{Junellie Perez}
\affiliation{Department of Earth \& Planetary Sciences, Johns Hopkins University, Baltimore, Maryland, USA}
\email{jgonza70@jhu.edu}

\author[0009-0009-5036-3049]{Emma Postolec}
\affiliation{Kapteyn Astronomical Institute, University of Groningen, P.O. Box 800, 9700 AV Groningen, The Netherlands}
\email{e.n.postolec@rug.nl}

\author[0009-0008-7799-7976]{Mariana Sastre}
\affiliation{Kapteyn Astronomical Institute, University of Groningen, P.O. Box 800, 9700 AV Groningen, The Netherlands}
\email{m.c.villamil.sastre@rug.nl}

\author[0000-0001-8106-6164]{Arnaud Salvador}
\affiliation{Institute of Space Research, German Aerospace Center (DLR), Rutherfordstr. 2, 12489 Berlin, Germany}
\affiliation{Université Paris Cité, Institut de physique du globe de Paris, CNRS, F-75005 Paris, France}
\email{salvador@ipgp.fr}

\author[0000-0002-5057-0322]{Hanno Spreeuw}
\affiliation{Netherlands eScience Center, Science Park 402, NL-1098 XH Amsterdam, The Netherlands}
\email{h.spreeuw@esciencecenter.nl}

\author[0000-0001-6974-6714]{Andrea Zorzi}
\affiliation{Stanford University, Department of Earth and Planetary Sciences, 450 Jane Stanford Way, Stanford, 94305, CA, USA}
\email{zorzi@stanford.edu}

\author[0000-0002-5967-9631]{Thomas J. Fauchez}
\affiliation{NASA Goddard Space Flight Center, 8800 Greenbelt Road, Greenbelt, MD 20771, USA}
\affiliation{Integrated Space Science and Technology Institute, Department of Physics, American University, Washington DC, USA}
\affiliation{NASA GSFC Sellers Exoplanet Environments Collaboration, USA}
\email{thomas.j.fauchez@nasa.gov}

\author{Keiko Hamano}
\affiliation{Division of Science, National Astronomical Observatory of Japan, 2-21-1 Osawa, Mitaka, Tokyo 181-8588, Japan}
\email{keiko.i.hamano@gmail.com}

\author[0000-0002-3555-480X]{Jérémy Leconte}
\affiliation{Laboratoire d’astrophysique de Bordeaux, Univ. Bordeaux, CNRS, B18N, allée Geoffroy Saint-Hilaire, 33615 Pessac, France}
\email{jeremy.leconte@u-bordeaux.fr}

\author[0000-0001-8804-120X]{Maxime Maurice}
\affiliation{Laboratoire Magmas et Volcans, CNRS, 63170 Aubière, France}
\email{maxime.maurice@uca.fr}

\author[0000-0001-8817-1653]{Lena Noack}
\affiliation{Department of Earth Sciences, Freie Universität Berlin, Malteserstrasse 74-100, 12249 Berlin, Germany}
\email{lena.noack@fu-berlin.de}

\author[0000-0002-5422-8794]{Laurent Soucasse}
\affiliation{Netherlands eScience Center, Science Park 402, NL-1098 XH Amsterdam, The Netherlands}
\affiliation{IMEC, Leuven, Belgium}
\email{Laurent.Soucasse@imec.be}

\begin{abstract}
Spectroscopic characterization of rocky exoplanets with the James Webb Space Telescope has brought the origin and evolution of their atmospheres into the focus of exoplanet science. Time-evolved models of the feedback between interior and atmosphere are critical to predict and interpret these observations and link them to the Solar System terrestrial planets. However, models differ in methodologies and input data, which can lead to significant differences in interpretation. In this paper, we present the experimental protocol of the Coupled atmospHere Interior modeL Intercomparison (CHILI) project. CHILI is an (exo-)planet model intercomparison project within the Climates Using Interactive Suites of Intercomparisons Nested for Exoplanet Studies (CUISINES) framework, which aims to support a diverse set of multi-model intercomparison projects in the exoplanet community. The present protocol includes the initial set of participating magma ocean models, divided into evolutionary and static models, and two types of test categories, one focused on Solar System planets (Earth \& Venus) and the other on exoplanets orbiting low-mass M-dwarfs. Both test categories aim to quantify the evolution of key markers of the links between planetary atmospheres and interiors over geological timescales. The proposed tests would allow us to quantify and compare the differences between coupled atmosphere--interior models used by the exoplanet and planetary science communities. Results from the proposed tests will be published in dedicated follow-up papers. To encourage the community to join this comparison effort and as an example, we present initial test results for the early Earth and TRAPPIST-1 b, conducted with models differing in the treatment of energy transport in the planetary interior and atmosphere, surface boundary layer, geochemistry, and the in- and outgassing of volatile compounds.
\end{abstract}


%

\section{Introduction} \label{sec:intro}
The study of exoplanets has entered a new era with the James Webb Space Telescope (JWST) already providing detailed observations of warm to hot rocky exoplanets \citep{Zieba2023, Hu2024, Gressier2024, Patel2024, Bennett2025, Espinoza2025, Teske2025}, and the Extremely Large Telescopes (ELTs) coming soon, we are now able to investigate the atmospheres, structures and compositions of rocky exoplanets in much greater detail. Because the atmospheres of these planets are a direct consequence of their interior evolution and outgassing, we need reliable interior models that are accurate, consistent, and comparable to each other in order to improve interpretation of atmospheric data.

Models that simulate the interior of planets (i.e., the condensed fraction that is not part of the gaseous envelope) are a key link between what we observe (mass, radius, and atmospheric composition) and what we want to understand, including a planet's composition, internal structure, and thermal evolution. However, resolving the connection between the interior and atmosphere is complex and the composition of the atmospheres depends on both the volatile content of the building blocks from which the planets formed and on how materials crystallize, differentiate, settle, and mix within their interiors over time \citep{Lichtenberg2023,Lichtenberg2025Science}. Different models often make different assumptions about these processes, leading to various estimates in magma ocean durations \citep[e.g., gray vs. non-gray atmosphere treatment;][]{Lebrun2013, Nikolaou2019} and resulting surface conditions. These differences affect our interpretations for the formation and evolution of planets, their ability to keep volatiles, and their potential to support habitable conditions.

Comparing and testing different models is therefore essential. Systematic multi-model studies and model intercomparison projects allow us to identify which assumptions and parameters have the largest impact on the results, and where uncertainties remain \citep{Fauchez21_workshop, Sohl24_cuisines}. It is now critical to undertake model intercomparisons to ensure that our theoretical capabilities match the pace of JWST and future ELT observations, which will continue to give us insight into the physics and nature of rocky exoplanets, providing multiple opportunities to connect models with real data \citep{Lichtenberg2025}. Once observations become available, the intercomparison framework will in turn allow efficient interpretation and feedback between modeling and observations—just as similar synergy has been advancing our understanding of formation scenarios of sub-Neptunes and super-Earths.

Improving the consistency and reliability of interior models also helps us decipher the early evolution of our own Solar System. Our understanding of the climatic and geological differences between Earth and Venus \citep{Hamano2013, Salvador2023_SSR}, the conditions on early Earth \citep{Miyazaki2022,Schaefer2024JGRE}, the past climate of Mars \citep{Schaefer2024JGRE,Sim2024JGRE}, and the reconstructed initial volatile inventories of the terrestrial planets \citep{Salvador2017,Schaefer2017ApJ} all depend on how we model their interiors and coupled interior-atmosphere interactions.

By systematically comparing a set of different models that aim to treat these effects, this work aims to build a stronger foundation for interpreting exoplanet observations, understanding planetary diversity, and assessing habitability and possible biosignatures.  

\subsection{\chili within CUISINES}

\chili is part of the CUISINES\footnote{\url{https://nexss.info/cuisines}} (Climates Using Interactive Suites of Intercomparisons Nested for Exoplanet Studies) framework \citep{Sohl24_cuisines}, which enables a standardized comparison of numerical models for exoplanets of various complexity and applications.
The first such model intercomparison within CUISINES was the TRAPPIST-1 Habitable Atmosphere Intercomparison \citep[THAI,][]{Fauchez20_thai_protocol} project, which focused on habitable climate scenarios for TRAPPIST-1\,e simulated by 3D general circulation models \citep{Turbet22_thai, Sergeev22_thai, Fauchez22_thai}.
THAI's success inspired nine additional and ongoing intercomparisons  \citep{Sohl24_cuisines}.
These projects encompass a range of exoplanets and Solar System planets, and the modeling of various physical processes such as radiative transfer \citep{Villanueva24_modeling}, atmospheric general circulation \citep{Christie22_camembert}, climate habitability and transitions \citep{Haqq-Misra22_samosa, Deitrick23_functionality}.

Here, we describe the \chili protocol, which specifies scenarios for modeling the evolution of several rocky planets: Earth, Venus and exoplanets in the TRAPPIST-1 system.
Thematically, \chili overlaps with THAI by including the TRAPPIST-1\,e case.
While THAI focused on potentially habitable, temperate climate scenarios, the models participating in the \chili project may inform the validity of such scenarios by providing theoretical predictions for the early composition of the secondary, outgassed atmosphere of TRAPPIST-1\,e.
\chili results may inform and motivate a potential follow-up intercomparison project to THAI.

Two ongoing intercomparison projects, PIE (Harman et al., in prep.) and COD ACCRA (Chaverot et al., in prep.), for 1-D photochemical and radiative-convective models, respectively, include a magma ocean planet case in their protocols.
This shared modeling regime thus provides a direct link from these ongoing intercomparisons to \chili.
In these two protocols, a magma ocean scenario is specified as an outgassed atmosphere consisting of rock vapors plus H$_2$, CO$_2$, H$_2$O volatiles, with the surface temperature set to $\approx$1,000\,K. However, \chili can provide physically-informed constraints on the expected atmospheric compositions overlying planetary magma oceans, and thus place the scenario assumed by PIE and COD ACCRA in the context of planetary evolution.
Reciprocally, results from PIE and COD ACCRA may be used to validate calculations of the energy fluxes and chemical composition in the atmospheric domain of the \chili models.
Additionally, radiative transfer models benchmarked in the MALBEC intercomparison project \citep{Villanueva24_modeling} will be able to generate synthetic transmission and emission spectra based on \chili output for the exoplanet cases. These synthetic spectra can then give a quantitative estimate of the detectability of potentially observable planetary magma oceans.
The standardized output files generated by \chili models (see Sec.~\ref{sec:output}) will ensure re-usability of the project's results by other CUISINES projects and the rest of the planetary science community. 

Following the CUISINES philosophy, \chili aims to maximize community participation through careful experimental design, separated from the paper(s) analyzing the results, and careful scheduling of the project timeline.
In this paper, we describe the formal \chili protocol to compare and benchmark multiple atmosphere-interior coupled models.
Participation in \chili \rev{remained} open until the end of winter 2025/2026\footnote{\rev{Exact dates can be found at the \href{https://github.com/projectcuisines/chili}{\chili GitHub repository}.}}.
A low entrance barrier across the experiments allows for broad participation, with no requirement for a given model to perform \textit{all} of the experiments.

\section{Motivations for \chili}\label{sec:motiv}

Due to the energetics of planetary accretion, all rocky planets most likely experience one or several magma ocean stages early in their history \citep[e.g.,][]{Tonks1993, Elkins-Tanton2012}. In this context, we refer to magma oceans as the early, transient stage in planetary formation and evolution, distinct from the steady state ``lava worlds", which are close-in planets with equilibrium temperatures high enough to produce silicate melting at the planet's surface and thus host a permanently (or episodically) molten surface due to stellar irradiation. The incoming stellar flux necessary to create a lava world is dependent on the atmospheric properties of the planet, and so no strict irradiation regime can be specified. Insights from this intercomparison will ultimately contribute to a unified understanding of rocky planet formation and evolution, bridging both transient and permanent magma ocean regimes.
The magma ocean stage of a planet's evolution is critical for determining its chemical differentiation, because the low viscosity of a vigorously convecting molten mantle and the absence of a stiff surface boundary layer allow for enhanced mixing between the gaseous envelope and planetary interior. A magma ocean period is thus likely responsible for the Earth's early outgassing from within its interior which built up the atmosphere \citep{Brown1952, Hamano1978, Sarda1985, Stuart2016}.

In turn, the optical thickness of a growing atmosphere modulates radiative cooling to space and the surface temperature, thereby feeding back on the interior's thermal evolution \citep[e.g.,][]{ElkinsTanton2008, Hamano2013, teller_early_1973}. This reciprocal exchange determines both the rate at which a planet cools and the mass and composition of its early atmosphere, establishing the initial volatile distribution and redox state that shape its subsequent climate and habitability.
Further, the solidified mantle volatile content and thermal state inherited from magma ocean solidification will control the post-magma ocean geodynamic regime and associated volcanic degassing \citep[e.g.,][]{Salvador2023_SSR,Lichtenberg2023}.

Within the \chili framework, a central objective is to understand which early scenarios may permit the later-formation of surface water oceans and, more generally, how surface conditions compatible with habitability emerge. As a magma ocean cools, volatile partitioning between melt, solid, and atmosphere sets the partial pressures of condensable species in the atmosphere. Once radiative cooling lowers the surface temperature below the saturation threshold for \ch{H2O}, condensation initiates and liquid water oceans can accumulate \citep[e.g.][]{Abe1985, Kasting1988, ElkinsTanton2008}. The onset and persistence of these water oceans depends on the evolving balance between interior heat flux, stellar irradiation \citep{Kasting1988, feulner_Thefaint_2012}, atmospheric opacity, and cloud feedbacks \citep{Turbet2021}, as well as the redox state that governs the relative abundances of atmospheric compounds, such as \ch{H2O}, \ch{CO2} \citep{Hamano2013, Lebrun2013, Salvador2017} and \ch{H2}, \ch{CO}, \ch{NH3}, \ch{CH4}, and \ch{SO2} \citep{Krissansen-Totton_2024,nicholls_redox_2024,nicholls_tidal_2025}.

More broadly, ``habitable conditions'' can be framed as the emergence of clement surface environments sustainable for water oceans -- involving moderate temperatures, stable pressures, and long-lived volatile inventories -- rather than a particular climate end state. These surface conditions emerge from tight coupling between planetary interiors and atmospheres: crystallization and degassing regulate the supply of volatiles \citep{hirschmann2022magma}; atmospheric escape and photochemistry sculpt their long-term retention \citep{Kasting1993, Wordsworth2013}; and the evolving energy balance \citep{Turbet2021,Boer2025ApJ} determines whether condensed phases become stable or persist. By comparing models across this chain of processes, \chili identifies which assumptions most strongly control water ocean condensation and the stabilization of temperate surface conditions after magma ocean solidification, clarifying both robust predictions and key uncertainties for rocky planetary evolution within and beyond our Solar System.

There is also a direct connection between magma ocean evolution models and exoplanet biosignatures. The super-luminous pre-main sequence of M-dwarfs may drive extreme rates of H$_2$O photolysis and hydrogen escape, causing O$_2$ buildup that could later mimic an inhabited planet with oxygen production via photosynthesis \citep{luger2015extreme, wordsworth_ABIOTICO_2014}. However, atmospheric O$_2$ may be sequestered in surface magma oceans \citep{schaefer2016, wordsworth2018redox, barth2021magma}; whether O$_2$ persists on geologic timescales depends on the co-evolution of early magma oceans, atmospheric escape, and planetary redox \citep{krissansen2022predictions, cherubim2025oxidation}. \chili provides an opportunity to test whether predictions of O$_2$ build-up or removal are consistent across models, improving our understanding of O$_2$ as a reliable biosignature for future life detection missions \citep{meadows2018exoplanet, krissansen2022understanding}.

There are no direct analogues to actively molten, magma ocean planets in the present-day Solar System. The rocky bodies have long since cooled and solidified, so the early stages when their mantles and atmospheres co-evolved can only be inferred indirectly from geochemical and isotopic records \citep{ElkinsTanton2008, Hamano2013, Lebrun2013, hirschmann2022magma}. However, observations of hot rocky exoplanets now provide real-time laboratories for studying these conditions \citep{Lichtenberg2025,Lichtenberg2025Science}. 

Strongly irradiated lava worlds can offer valuable insights for temperate rocky worlds and their early magma ocean episodes, and are a uniquely promising target group for observation. If a lava world is depleted in volatile elements by a combination of formation and atmospheric escape, its atmosphere would be rock-vapor dominated. Such atmospheres provide a unique window to the mantle composition of rocky worlds, their chemistry \citep{vanBuchem2023, Piette2023ApJ} and dynamics \citep{Meier2023A&A, Herath2024MNRAS}. Recent observations indicate that some lava worlds such as 55~Cancri\,e and TOI~561 b \citep{Hu2024, Teske2025} likely maintain volatile-rich atmospheres that efficiently redistribute heat. These worlds may thus host permanent magma oceans, so characterizing them and interpreting their geophysical and climate states using comprehensive modeling can help constrain these planets' initial volatile inventory, formation scenario, atmospheric escape history and mantle geochemistry. 
    
Thanks to their high thermal brightness and short periods, lava worlds are ideal targets for JWST transit spectroscopy and phase curve measurements. Static atmospheric structure and chemistry models have recently been constructed for modeling these worlds \citep[e.g][]{vanBuchem2023, vanBuchem2025, peng2024, nguyen2022impact, nguyen2024clouds, cmiel2025, Piette2023ApJ} to interpret measurements of their masses, radii and spectroscopic characteristics. Meanwhile, modeling the evolution of these worlds offers a permanently molten end-member benchmark for the magma ocean evolution models which are otherwise calibrated on Earth-like mineralogy and volatile chemistry. Conducting model intercomparison in this highly irradiated regime identifies assumptions and limitations among the static models, as well as the applicability of evolutionary models in interpreting lava world observations. Additionally, future imaging surveys of young exoplanetary systems may capture these transient magma ocean phases directly, which would open another window into the interaction between magma oceans and their outgassed atmospheres, distinct from lava worlds \citep{Lupu2014ApJ,Hamano2015,Bonati2019A&A,Cesario2024A&A}.

\section{Description of the participating models} \label{sec:models}

In this section we provide baseline descriptions and references of the codes that we anticipate will participate in the \chili intercomparison. Two classes of models will participate: (i) evolutionary models and (ii) static models. Evolutionary models aim to resolve the time evolution sequence of how planets cool from an initially hot and molten state, crystallize their mantles, and outgas their atmospheres. In comparison, static models provide a snapshot in time with fixed boundary conditions. In the following sub-sections, we describe the participating codes and their main assumptions, specifically highlighting overlaps and differences between codes. These descriptions are not exhaustive, but serve as summary descriptions for the main workings of each code and highlight their differences. 

\begin{figure*}[ht]
    \centering
    \includegraphics[width=\linewidth]{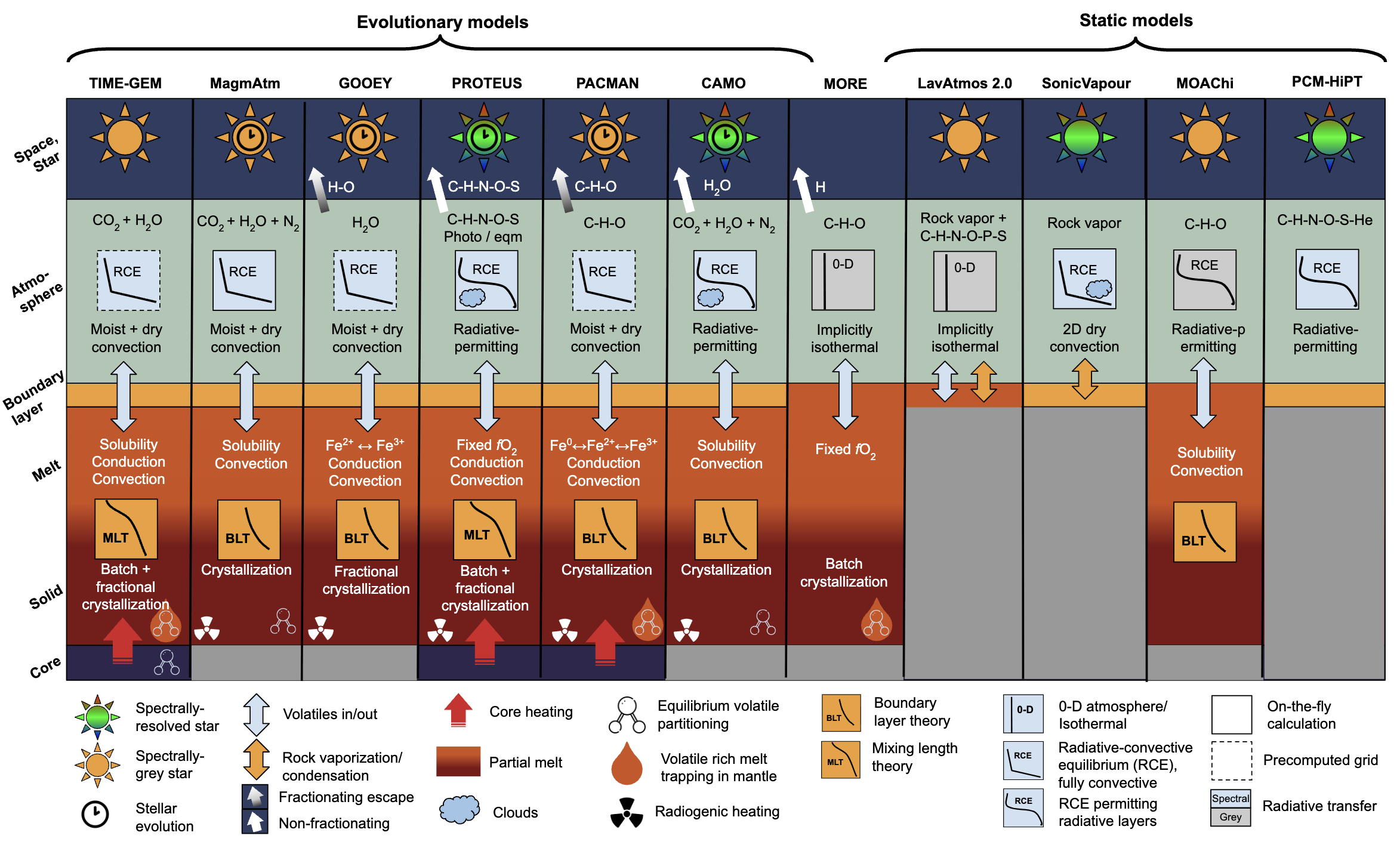}
    \caption{Illustration describing the processes treated by several models participating in \chili. The leftmost seven models are evolutionary, where dynamics and composition change with time. The rightmost four models are static, where the planets  are implicitly assumed to be at steady-state. MORE is shown between evolutionary and static models, as it not fully time-dependent (see main text). The six radial domains of the models include: the core, solid mantle, liquid mantle, surface boundary layer, atmosphere, and the exosphere or outer-space. Interactions coupling these layers are denoted by arrows.}
    \label{fig:overview_models}
\end{figure*}

A schematic overview of the participating codes is displayed in Fig.~\ref{fig:overview_models}, which illustrates the main physical processes involved and their  prescriptions within the models. Points of contact and primary literature references may be found in Table~\ref{tab:models}. Many of these models are under active development; we describe their state at the time of submission.
\rev{For the intercomparison, all models adopt a prescribed constant Bond albedo of 0.1, regardless of their native albedo treatment capabilities, to isolate differences in interior--atmosphere coupling from surface reflectivity assumptions.}

There are modeling regimes which sit outside the scope of all participating models. None of these models spatially resolve the planet's metallic core, although may treat its properties and behavior in bulk. It has also been suggested that larger sub-Neptune planets may not have distinct boundaries between their atmospheres, mantles, and cores, due to physical or chemical mixing between these phases \citep{Lichtenberg2021ApJL, Pierrehumbert2023ApJ, Vazan2024AA, benneke_toi270d_2024, young_subneptune_2025}. However, all of the \chili models assume distinct core-mantle and mantle-atmosphere boundaries, adopting various boundary-layer treatments at these interfaces. \rev{Core--mantle interaction (e.g., parameterized core cooling or core heat flux) is not required for the baseline intercomparison cases; models that include it should document their treatment in the accompanying notes file. The extent to which participating models treat core--mantle coupling is partly illustrated by the red upward arrows in Fig.~\ref{fig:overview_models}.} The `atmosphere' is equivalent to the planet's volatile outer `envelope'. 

\begin{table*}
\caption{Coupled atmosphere-interior models in \chili, their point of contact, and primary model references.}
\begin{tabular}{l l l}
\hline
Code & Contact & Primary references \\
\hline
TIME-GEM & Y.~Miyazaki & \cite{Miyazaki2022} \\
MagmAtm & A.~Salvador & \cite{Salvador_MagmAtm} \\
GOOEY & L.~Schaefer & \cite{schaefer2016} \\
PROTEUS & T.~Lichtenberg, H.~Nicholls
& \cite{lichtenberg2021, nicholls_redox_2024, nicholls_convective_2025, nicholls_tidal_2025} \\
PACMAN & J.~Krissansen-Totton & \cite{Krissansen-Totton_2024} \\
CAMO & H.~Pelissard, J.~Leconte & \emph{Pelissard et al. In Prep} \\
MORE & L.~Noack, P.~Baumeister & \emph{Noack et al. In Prep} \\
LavAtmos~2.0 & Y.~Miguel, L.~Janssen & \cite{vanBuchem2023, vanBuchem2025} \\
MOAChi & B.~Peng & \cite{peng2024} \\
SonicVapour & T.~G.~Nguyen & \cite{nguyen2022impact,nguyen2024clouds} \\
PCM-HiPT & J.~Cmiel & \cite{cmiel2025} \\ 
\hline
\label{tab:models}
\end{tabular}
\end{table*}

\subsection{Evolutionary models}

\subsubsection{TIME-GEM} \label{timegem}
TIME-GEM (Terrestrial Interior Magma Evolution with Gibbs Energy Minimization) is a 1-D coupled thermochemical evolution model for a planet's magma ocean and atmosphere \citep{Miyazaki2022}, focusing on a physically self-consistent treatment of volatile sequestration within the solidifying mantle. The model calculates temperature evolution using convective scaling and mixing length theory. The solidification state and the stable phase assemblage are predicted via Gibbs free energy minimization \citep{Miyazaki2019} at each depth and timestep. The phase-equilibria calculation, coupled with the subsequent modeling of melt–solid separation and volatile migration through two-phase flow, predicts volatile transport within the solidifying magma ocean. This characterizes the volatile budget of the magma ocean and the resultant degassing flux to the CO$_2$-H$_2$O atmosphere. The atmospheric component uses a 1-D radiative–convective structure, with the outgoing longwave radiation computed by petitRADTRANS, and is assumed to remain in thermochemical equilibrium with the surface magma ocean throughout the evolution.

\subsubsection{MagmAtm} \label{magmatm}
MagmAtm is a coupled magma ocean--atmosphere thermal evolution model designed to capture the solidification and outgassing of a terrestrial planet's last magma ocean stage and its transition into solid, quasi steady-state evolution (\citealp{Salvador_MagmAtm}; inherited from \citealp{Salvador2017}). The radiative-convective (moist and/or dry convection) atmosphere is comprised of \ch{H2O}, \ch{CO2} and \ch{N2}, and radiative transfer is enforced by either a gray or $k-$correlated scheme \citep{Marcq2012, Marcq2017}. The thermal evolution is driven by radiogenic heating, mantle convection, and radiative cooling. Volatiles are partitioned between the magma and the atmosphere through solubility laws, and between the melt and crystals through partition coefficients.

\subsubsection{GOOEY} \label{schaefer}

The newly-named GOOEY code \citep{schaefer2016} is a coupled magma ocean-atmosphere model designed to track the evolution of oxygen (\ch{O2}) and water (\ch{H2O}) in a steam atmosphere undergoing hydrodynamic escape. It allows an evolving oxygen fugacity through the interaction of \ch{O2} and iron oxides in the magma ocean. The model includes secular cooling of the interior, long-lived radionuclides, and latent heat of crystallization as heat sources. The model adopts a thermal boundary layer model that includes convection within the mantle interior, which is assumed to follow an adiabatic temperature profile, and conduction through a surface boundary layer that controls the heat flux into the bottom of the atmosphere. This model does not take into account melt-trapping during crystallization, but does include volatile partitioning into the solid mineral phase. The mantle viscosity is dependent on melt fraction but not volatile content. GOOEY does not account for density evolution in either the solid or liquid layer. The atmosphere adopts pre-computed solutions to outgoing longwave radiation from a line-by-line (LBL) radiative-convective model. Energy-limited escape is included, which allows fractionation between hydrogen and oxygen. Lastly, it accounts for stellar evolution for the XUV fraction that drives escape, but does not adopt a specific stellar spectrum. 

\subsubsection{PROTEUS} \label{proteus}
PROTEUS is a modular numerical framework that simulates the 1-D mantle and atmospheric thermo-compositional evolution of rocky planets \citep{lichtenberg2021,nicholls_redox_2024}. PROTEUS tracks evolution by time-stepping the mantle dynamics according to mixing-length convection, gravitational settling, latent heating, and conduction \citep{bower_numerical_2018}. The mantle vertically resolves rheological properties as a function of melt fraction, including two-phase flow and parameterised batch and fractional crystallization. Internal heat sources include time-dependent radionuclide decay, parametrised metallic-core cooling, and tidal dissipation. PROTEUS accounts for the partitioning of C-H-N-O-S volatiles between the atmosphere and melt, assuming a compositionally homogeneous mantle and a constant oxygen fugacity fixed relative to the iron-w\"ustite buffer. Volatile elements speciate in the atmosphere through 1-D equilibrium thermochemistry. The atmosphere is solved for radiative–convective equilibrium at each interior time-step, such that energy fluxes are conserved on a local and global basis, and the formation of stable radiative layers is permitted \citep{nicholls_convective_2025}. Non-fractionating hydrodynamic mass loss is modeled in the energy-limited regime. Stellar properties, including spectroscopic fluxes, are evolved self-consistently alongside the planet throughout the simulation \citep{johnstone2021active}. Hydrostatic density structure is evaluated through the planet: held fixed within the interior, and updated self-consistently within the atmosphere.

\subsubsection{PACMAN} \label{pacman}
The Planetary Atmosphere, Crust, and MANtle geochemical evolution model (PACMAN) captures the coupled thermal, atmospheric, and redox evolution of a solidifying magma ocean \citep{krissansen-totton2021, krissansen2021venus}. The version of the model considered here, PACMAN-P, accommodates arbitrary \ch{CO2}-\ch{H2O}-\ch{H2}-\ch{CO}-\ch{CH4}-\ch{O2} atmospheric compositions, and includes melt solubilities for \ch{H2O}, \ch{CO2}, \ch{H2}, and graphite, as well as iron speciation in the solid and molten mantle \citep{Krissansen-Totton_2024}. The oxygen fugacity of the solid and molten mantle co-evolve with the solidifying magma ocean. Surface climate is calculated by assuming a dry adiabat-moist adiabat-isothermal temperature profile and computing outgoing longwave radiation using petitRADTRANS, which uses correlated-k opacities \citep{Mollière_2019}. Internal heat sources include primordial heat, radiogenics, specified metallic core heating, and the latent heat of solidification, and interior heat flow is determined by convection scaling laws, along with temperature dependent mantle viscosity. At every timestep, the model solves a coupled system of equations that ensure chemical equilibrium between the atmosphere and dissolved species (including iron species), the oxygen fugacity of the melt, and thermal equilibrium between absorbed stellar flux, heatflow from the interior, and outgoing longwave radiation. Atmospheric escape is either diffusion-limited or XUV-limited depending on composition/stellar fluxes, and in the XUV-limited regime C- and O-bearing species can be dragged along in the hydrodynamic wind, resulting in fractionating escape.  As the solidification front moves from the core-mantle boundary to the surface, both equilibrium partitioning and melt trapping transfer volatiles from the melt to the solid mantle. 

\subsubsection{CAMO} \label{camo}
CAMO (Coupling Atmosphere–Magma Ocean) aims to couple a thermal evolution model of rocky planets with a 0-D internal model \citep{Lebrun2013, Salvador2017} and the self-consistent 1-D radiative–convective atmospheric model Exo\_k \citep{leconte_spectral_2021}. The interior module provides the time-dependent internal heat flux and volatile outgassing (H$_2$O/CO$_2$), while Exo\_k returns the corresponding equilibrium surface temperature. At each interior timestep, CAMO assumes a quasi-static atmospheric response and iteratively solves for the atmospheric equilibrium state consistent with the evolving internal conditions. Radiative transfer is computed using correlated-k or line-by-line (LBL) opacities, with dry and moist convective adjustments, condensation/evaporation, and radiative layers treated to maintain local energy balance. At every step, the model solves the surface energy balance by root-finding to jointly satisfy atmospheric equilibrium and internal convection, determining the adaptive timestep. The framework supports variable insolation, stellar spectra, and tracks surface-pressure growth from degassing and atmospheric escape.

\subsubsection{MORE} \label{more}
MORE is a simplified Magma Ocean model including Redox Evolution of the atmosphere and is based on the solidification models of intrusive magmas \citep{vulpius2022intrusive}, scaled to global magma oceans. The model does not calculate the thermal cooling of the planet but instead assumes that continuous crystallization occurs via a batch crystallization model. There is therefore no actual time-dependence in the model. For this reason, Fig. \ref{fig:overview_models} shows MORE in between the evolutionary and static models. Partition coefficients together with melt trapping via a free trapping efficiency factor are used to track the partitioning of volatiles between solid mantle and magma ocean. Volatiles are expected to be homogeneously mixed in the magma ocean which is at chemical equilibrium with the atmosphere. Outgassing and ingassing of multiple C-H-O species are calculated via Henrian solubility laws, redox-dependent gas speciation models for gas-melt equilibrium \citep{brachmann2025distinct}, and re-equilibration of the atmosphere via FastChem \citep{kitzmann2024fastchem}. Atmospheric escape is not directly calculated, but efficient loss of H$_2$ can be considered, leading to complete desiccation of the magma ocean. The MORE model is intended to be a very fast and easy-to-couple, first-order approximation of the primary outgassed atmosphere and interior volatile storage, which is intended to be directly coupled to long-term evolution codes. This has already been done for the TEMPURA code \citep{Baumeister2023}, a 1-D coupled atmosphere-interior model of the long-term (solid-state) evolution of rocky planets, which takes the end-state temperature, atmosphere, and volatile content in the mantle calculated by MORE as a starting point for the planet's long-term evolution.

\subsection{Static models}
Static models describe the state of a planet at a specific point in time without accounting for its evolution. These models are generally developed to explain exoplanet observations obtained at a given stage in a planet’s history, typically focused on lava planets, as defined in the introduction. Because they are simplified, they run quickly and can perform a large number of calculations, making them well suited for retrieval analyses that aim to interpret observations \citep[e.g.,][]{zilinskas2025}.

\subsubsection{LavAtmos 2.0} \label{lavatmos}

LavAtmos is an open-source code originally developed to study lava worlds with silicate vapor atmospheres. It calculates the equilibrium between a planet’s molten surface and its atmosphere, estimating the partial pressures of gases that are vaporized from the magma and remain in equilibrium with the overlying atmosphere \citep{vanBuchem2023, vanBuchem2025}. The code takes as input the magma composition, its temperature (as determined by stellar irradiation), and the initial atmospheric composition, and then computes the abundances of the gases vaporized from the magma that are in equilibrium with the atmosphere. The oxygen fugacity of both the magma and the atmosphere is calculated self-consistently, ensuring that the system satisfies the laws of mass action and mass conservation in a mixture containing both melt species and volatile elements.
The original version of the code considered pure rock vapor atmospheres \citep{vanBuchem2023}. The updated version, LavAtmos 2.0, expands these capabilities by including both rock vapor and volatile-rich atmospheres containing C-H-N-O-S-P-bearing species. In this new version, the equilibrium between the molten surface and the atmosphere is computed using a coupled approach with the FastChem code \citep{Stock2018}, allowing the treatment of 532 chemical species in total.

\subsubsection{SonicVapour} \label{sonicvapour}

SonicVapour simulates the hydrodynamics of collapsible, silicate vapour atmospheres on lava worlds \citep{nguyen2024clouds}. It determines the steady-state atmospheric pressure, temperature, wind speed, and evaporative rate with respect to angular distance from the substellar point. The atmosphere model calculates 1-D LBL radiative transport on prescribed adiabatic or isothermal $P$-$T$ profiles, accounting for the variation in the phase angle from the substellar point, assuming rotational symmetry. Atmospheric composition is calculated from equilibrium chemistry models such as LavAtmos with user-specified mantle composition (typically Bulk Silicate Earth composition). Spectral analogues from the MUSCLES survey are used to extract analogous stellar spectrum \citep{france2016muscles}. Silicate cloud formation can be inferred from calculated condensation rates.

\subsubsection{MOAChi} \label{moachi}
The Magma Ocean - Atmosphere Chemistry (MOAChi) model is a coupled atmosphere - interior structure model for lava worlds with thick C-O-H atmospheres \citep{peng2024}. The model estimates the transit height of an atmosphere given the planet's rocky mass, equilibrium temperature and total (atmosphere + magma ocean) C and H inventories. It includes a 1-D radiative-convective atmosphere using a dual grey scheme, allowing for the convective shutdown in deep atmosphere \citep{Guillot_2010}. The atmosphere includes 6-species, C-H-O chemical equilibrium with non-ideal equation of state  \citep[EoS, ][]{Duan_1992_EOS,Duan_1996_EOS}. The user can input a C/O ratio for the atmosphere as an observable proxy for the redox state of the system. The mantle module assumes an adiabatic profile using the lower mantle EoS from \cite{Valencia_2006, Plotnykov_Valencia_20_cmf}, and calculates the magma ocean depth and dissolved volatile abundances using the surface temperature and Henrian solubilities. Numerically, the model iterates over top-of-atmosphere locations and atmospheric C/H to partition a user-specified bulk C and H endowment between the atmosphere and the magma ocean, and outputs the atmospheric profile and transit radius solutions.

\subsubsection{PCM-HiPT} \label{pcm}
Planetary Climate Model for High Pressures and Temperatures (PCM-HiPT) is a one-dimensional, line-by-line radiative–convective model designed to simulate the thermal structure of dense, hot terrestrial exoplanet atmospheres \citep{cmiel2025}. It employs a high-resolution spectral grid and HITRAN-based absorption data \citep{Gordon2022} to model radiative energy transfer with high accuracy at elevated pressures and temperatures ($>1000$ K). Within the \chili intercomparison framework, PCM-HiPT serves as the atmospheric radiative–convective solver. It can be coupled to interior models at each physical timestep by providing an equilibrium atmospheric state, enabling fully time-dependent coupling between surface and atmospheric evolution. The model computes temperature–pressure profiles, outgoing longwave radiation, and vertical thermal structure, including possible radiative inversions or decoupling layers. It maintains numerical stability and radiative–convective consistency under extreme pressure–temperature conditions relevant to magma ocean and post-runaway regimes.

\section{The \chili Protocol and Experiments} \label{sec:protocol}

The following section discusses the physical and chemical aspects of the intercomparison, and the input and output parameters and their motivation for the \chili project. Section \ref{sec:output} describes the necessary and optional model output quantities and their technical definitions. The \chili GitHub repository\footnote{\url{https://github.com/projectcuisines/chili}} will serve as a dynamic repository for input configurations and output data, which will continue to evolve during the intercomparison implementation. All subsequent repository versions are permanently archived on the \rev{\href{https://doi.org/10.5281/zenodo.17312304}{CHILI Zenodo Archive}} \citep{chili_zenodo}. We refer the reader to the GitHub repository in the case of technical updates following this article. All model outputs are specified in detail again in this repository and their format can be copy-pasted from other model output for simplicity.

\subsection{Earth and Venus}
\label{sec:solarsystem}
Coupled magma ocean-atmosphere models were originally developed to understand the early evolution of Earth and Venus \citep{Abe1985, Kasting1988}, and so \chili includes several Earth/Venus cases to compare outcomes for planets for which we have in-situ constraints on their thermal state. To enable a comparison against these constraints, we specify a small set of scenarios for Earth and Venus for which models are run. We assume identical initial conditions for both of these planets (Table \ref{tab:inputs_solar_system}) with the exception of their instellation, mass, and radius (using $R_\oplus=6.3781\times10^6~\rm m$ as the nominal Earth radius; \citealp{Prsa2016}). In Table~\ref{tab:inputs_solar_system}, we provide values for the relevant input parameters, as well as a representative values for potential initial H and C inventories with which to perform a grid search against the constraints. Both planets are treated as non-synchronously rotating. The simulations will begin from a fully molten state, neglect the radiative effects of clouds, and assume a constant planetary Bond albedo of 0.1 with a fixed zenith angle following \citet{Hamano2015}. 

\rev{Unless otherwise stated, models adopt no radiogenic heating for the baseline intercomparison cases. Models that include radiogenic heating should specify their adopted heat production rates and isotopic systems in the accompanying notes file.}

Inputs quantities include solar properties and planetary parameters. For the Sun, models that account for the luminosity evolution should optimally adopt a 1 solar-mass luminosity track from \citet{Baraffe2015} and construct the solar spectrum using blackbody radiation. Planet models should begin at a solar age of 50\,Myr. Those models that adopt a constant luminosity should use a Faint Young Sun model, with a reduced solar flux equal to 920\,$\rm W\,m^{-2}$ at a distance of 1 AU (i.e., 67.6\% of the current flux, using the nominal solar conversion constant $S_0=1361~\rm W\, m^{-2}$ to compute the total solar irradiance; \citealp{Prsa2016}). For both planets, the composition of Bulk Silicate Earth (BSE) and an Earth-like core radius ($0.55 R_\oplus$) should be adopted. The initial volatile content is specified in terms of H and C masses (or H$_2$O and CO$_2$ equivalents). All of these quantities are included in Table \ref{tab:inputs_solar_system}.

To enable model intercomparison of evolutionary models, the following quantities should be reported in CSV-formatted text files with rows tabulating time-evolution: surface temperature ($T_s$), mantle potential temperature ($T_p$), average volumetric melt fraction ($\phi$), absorbed stellar radiation flux ($F_{\rm ASR}$), outgoing longwave radiation flux ($F_{\rm OLR}$), and surface geothermal heat flux ($F_{\rm surf}$), mantle oxygen fugacity ($f$\ch{O2}) in the melt and solid phases at the surface (bar), total masses of C and H in solid/melt/atmospheric phases (kg), surface partial pressures of atmospheric constituents (bar), atmospheric mean molecular weight ($\mu$), optical transit radius (m), solidification radius\footnote{Defined as the radius within the mantle at which the model-defined critical melt fraction is crossed, in a bottom-up crystallization scenario.} (m), mantle viscosity (Pa\,s), and the surface conductive thermal boundary layer thickness (m). If a given model does not compute one of these quantities, its corresponding column should be \rev{filled with NaN}. \rev{The oxygen fugacity ($f$\ch{O2}) is reported separately for the solid mantle and melt phases, as these may diverge during crystallization. The mantle viscosity refers to a characteristic whole-mantle effective value, representative of the bulk rheological state at each timestep.} Important points which clarify deviations or exceptions compared to the other models should be stored as an accompanying note (see Sect.~\ref{sec:output}).

Model comparison will be undertaken through two parallel approaches. Firstly, we will compare the complete time evolution of all output variables (Table~\ref{tab:out}), focusing on differences at pre-defined points in time (e.g. 10$^4$ years, 10$^5$ years) to enable a comparison between evolutionary and static models. Additionally, we will align modeled evolutionary trajectories so that simulations are compared at equivalent points in magma ocean solidification phase space; for example, a comparison is made when the magma ocean is 50\% crystallized, or surface temperature is 1000\,K, whenever that occurs in the time domain. 

Following this, inputs for static models will then be set from output quantities of the evolutionary models which predict the extrema of magma ocean solidification times. For this study, we allow individual models to adopt their natural definition of solidification time (e.g. threshold melt fraction, surface temperature below solidus) which should be stated in the explanatory notes for each model. Static models will be compared to evolutionary models at specific timesteps, as described in Section \ref{sec:staticmodels}.

For Earth, geological and geochemical measurements can place constraints on the physicality of simulation results. For instance, the presence of zircons suggests a solidified surface emerged on Earth within 200\,Myr from its formation \citep{Wilde2001, Valley2014}. Therefore, we anticipate that models will produce a surface temperature that falls between the solidus (1400\,K at the surface, e.g., \citealp{Andrault2011}) and equilibrium temperature of the planet given a plausible initial volatile inventory of volatiles. Furthermore, the presence of surface liquid water is suggested around the same time, requiring a surface temperature below the dew-point temperature of \ch{H2O} \citep{Wilde2001,Mojzsis2001Natur}. This constraint--along with estimates of modern BSE volatile inventories--enables an intercomparison of the predicted initial volatile inventories for the Earth which result in models consistent with these geologically-constrained temperature regimes. Similarly, for Venus, models should ideally predict a solidified surface at simulation times equivalent to the present day.

Below, we provide details on the initial conditions we set for Earth and Venus cases. Additionally, we describe the parameter ranges for grid search test cases for Earth. All simulations are to be run until magma ocean solidification (as defined naturally for each model) or for a maximum of 1\,Gyr if solidification does not occur.

\paragraph{Case 1: Nominal Earth}
For this case, the planet is to be initialized with 3 Earth ocean equivalents of water in the mantle-atmosphere system, or $4.2\times10^{21}$ kg H$_2$O (or equivalently $4.7\times10^{20}$ kg H) and $1\times10^{21}$ kg CO$_2$ (or $2.72\times10^{20}$ kg C). For models that independently specify initial oxygen or oxygen fugacity, these quantities should be chosen to recover a final mantle redox state of IW+4 (i.e., $f$\ch{O2} at 4 log-units above the iron-w\"ustite buffer\rev{, approximately representative of Earth's present-day upper mantle oxidation state}, at a given temperature) when magma ocean solidification is complete (assuming no escape). This means initial oxygen fugacity (or equivalently initial Fe$^{3+}$/Fe$_T$) should be set to ensure an oxygen fugacity of IW+4 in the residual magma ocean at the end of the simulation run (the definition of which is model-dependent). \rev{To enforce this value, models that fractionate Fe$^{3+}$/Fe$_T$ should adjust the initial Fe$^{3+}$/Fe$_T$ so that the $f$\ch{O2} computed from the final (fractionated) magma ocean's Fe$^{3+}$/Fe$_T$ and a temperature of 1600\,K using the redox model of \citet{hirschmann2022magma}, or equivalent redox parameterization, is IW+4.} The models which target a final oxygen fugacity of IW+4 should then re-run simulations with the identified initial redox state, but then accounting for all relevant physical processes -- including escape.

\paragraph{Case 2: Nominal Venus}
This case is identical to Earth (Case 1) except that planetary mass and radius are adjusted to match Venus, and the early solar luminosity is scaled to represent Venus' insolation at 0.723\,AU (1760\,W\,m$^2$). The initial $f$O$_2$ should be identical to the Earth case.
\rev{Present-day Venus provides key boundary constraints for model validation: a surface temperature of $T_{\rm surf} \approx 733$\,K, a surface pressure of $\sim$92\,bar dominated by \ch{CO2}, and trace \ch{H2O} at $\sim$30\,ppmv \citep{Taylor2018SSRv}. The median surface age of Venus is estimated at $\sim$0.5\,Gyr \citep{McKinnon1997}, suggesting relatively recent global resurfacing.}

\paragraph{Case 3: Earth grid}
For this case, each model is to be run across a grid of initial volatile inventories to explore the extent to which different models agree on the initial conditions consistent with in-situ constraints on Earth's thermal state. The planet is to be initialized with 1, 5, and 10 initial Earth oceans of H$_2$O (or equivalent masses of H), and 0.5$\times$10$^{21}$, 1$\times$10$^{21}$, and 2$\times$10$^{21}$ kg CO$_2$ (or equivalent mass of C), for a total of 9 model runs. More finely resolved initial volatile inventory grids may be submitted if computationally feasible. Initial oxygen fugacity (or mantle redox state) is to be specified the same as in Case 1.

\begin{deluxetable*}{lccc}
\tabletypesize{\scriptsize}
\tablewidth{0pt}
\tablecaption{Initial parameters for nominal Earth and Venus\label{table:out}}
\tablehead{
\colhead{Parameter} & \colhead{Earth}&\colhead{Venus} & \colhead{Grid samples}
}
\startdata
Starting simulation time [Myr] & 50 & 50 & -- \\
Stellar mass [$M_\odot$] & 1.0 & 1.0 & -- \\
Orbital period [days] & 365.26 & 224.7 & -- \\
Eccentricity & 0 & 0 & -- \\
Semi-major axis [AU] & 1 & 0.723 & --\\
Stellar irradiance [W.m$^{-2}$] & 920 & 1760 & -- \\
Bond albedo & 0.1 & 0.1 & -- \\
Planet mass [M$_\oplus$]& 1 & 0.815 & -- \\
Planet radius [R$_\oplus$] & 1 & 0.95&--\\
Core radius [fraction] & 0.55 & 0.55 & --  \\
T$_{\mathrm{mantle}}^{\mathrm{init}}$ & fully molten & fully molten & -- \\
Mantle composition  & BSE & BSE & -- \\
Initial oxygen fugacity at 1600 K* & IW+4 & IW+4 & -- \\
H [kg] / H$_2$O [kg]  & $4.7\times10^{20}$ /  $4.2\times10^{21}$ & $4.7\times10^{20}$ /  $4.2\times10^{21}$ &  [1.6, 7.8, 16] $\times~10^{20}$ / [1.4, 7.5, 14] $\times~10^{21}$\\
C [kg] / CO$_{2}$ [kg] & $2.73\times10^{20}$ / $1\times10^{21}$ & $2.73\times10^{20}$ / $1\times10^{21}$ & [1.36, 2.73, 5.44] $\times~10^{20}$ / [0.5, 1, 2] $\times~10^{21}$ \\
\enddata
\tablecomments{*Users should ensure that final mantle oxygen fugacity is equivalent to the IW+4 buffer.}
\label{tab:inputs_solar_system}
\end{deluxetable*}

\begin{table*}
\caption{Scalar output variables required from participating evolutionary models at each step of their time-integration. Other calculated variables are free to be recorded alongside these.}
\begin{tabular}{l l l}
\hline\hline
Output variable & Description & Units \\
\hline
$t$    & Time relative to initialization          & yr    \\
$T_\text{surf}$  & Surface mantle-atmosphere temperature    & K     \\
$T_\text{pot}$  & Effective mantle potential temperature   & K     \\
$F_\text{surf}$ & Surface geothermal heat flux             & $\rm W\,m^{-2}$ \\
$F_\text{OLR}$ & Outgoing longwave radiation (OLR)        & $\rm W\,m^{-2}$ \\
$F_\text{ASR}$ & Absorbed stellar radiation (ASR)         & $\rm W\,m^{-2}$ \\
$\Phi_v$ & Volumetric whole-mantle melt fraction    & --    \\
$f\ch{O2}$ & Oxygen fugacity of solid mantle \& melt  & bar, bar    \\
$d_\text{CBL}$ & Thickness of surface conductive boundary layer  & m    \\
$m_C$ & Mass of carbon atoms in atmosphere, solid, and melt & kg, kg, kg    \\
$m_H$ & Mass of hydrogen atoms in atmosphere, solid, and melt & kg, kg, kg    \\
$m_O$ & Mass of oxygen atoms in the atmosphere   & kg   \\
$p_\text{surf}$ & Total surface pressure   & bar   \\
$p_x$ & Partial pressure \ch{H2O}, \ch{CO2}, \ch{CO}, \ch{H2}, \ch{CH4}, \ch{O2}   & bar $\times6$  \\
$\mu$ & Mean molecular weight of the bulk atmosphere & $\rm kg\,mol^{-1}$    \\

$R_\text{trans}$ & Effective photospheric-transit radius    & m   \\
$R_\text{solid}$  & Radius of solidification/rheological front    & m   \\
$\nu$ & Characteristic viscosity of the whole mantle    & Pa\,s   \\
\hline
\label{tab:out}
\end{tabular}
\end{table*}

\rev{Evolutionary models should record} vertical atmospheric profiles of temperature (K), pressure (bar), composition (volume mixing ratio), and height relative to the surface (m) in CSV-formatted text files. These profiles should be tabulated for each planet at the simulated times closest to those defined by Table \ref{tab:snapshots_times}, thereby enabling a direct comparison between the atmospheric calculations derived from evolutionary models and those from static models. Temperature-pressure profiles should be reported for Earth, Venus, and each exoplanet case. \rev{This is a recommendation rather than a strict requirement, as not all evolutionary models spatially resolve the atmosphere.} \rev{Opacity profiles are welcomed as optional auxiliary output where available.}

\subsection{Exoplanets}
\label{sec:exoplanets}

In addition to comparing the participating models under the Earth and Venus scenarios, we test their differences for exoplanet studies. We define three cases for these exoplanet intercomparison studies: an Earth-sized planet with the orbital parameters of TRAPPIST-1\,b, e and $\alpha$. TRAPPIST-1\,$\alpha$ is a hypothetical planet orbiting TRAPPIST-1 with a semi-major axis of $6.75\times 10^{-4}$\,AU, corresponding to an orbital period of 0.02136\,days and a radiative equilibrium temperature of $\sim1600$\,K. We note that this configuration implies an orbital separation within the stellar Roche limit of TRAPPIST-1, and is therefore not representative of any physically stable orbit. Nevertheless, this configuration serves as an important proxy for the extreme radiative forcing relevant to magma ocean conditions, allowing us to test the models' responses in a high-temperature limit that approximates early planetary evolution, later-stage accretion environments, and the irradiation conditions of some ultra-short period exoplanets around larger-mass stars (e.g. 55 Cancri\,e, TOI-561\,b). We adopt TRAPPIST-1\,$\alpha$ to enable us to test these irradiation conditions without varying additional system parameters that would introduce substantial complexity to \chili. Rather than using literature-derived planetary properties for TRAPPIST-1\,b and e, we assume Earth-like values for the mass and radius for all three of these planets in order to highlight differences between model frameworks, and to limit the propagation of observational uncertainties. \rev{This deliberate design choice isolates model framework differences from input parameter variations, enabling a cleaner comparison of how each code treats the same physical problem.} \rev{TRAPPIST-1\,$\alpha$ thus serves to test how well models handle the transition between a transient magma ocean planet and a permanent lava world.}
We adopt the stellar parameters of TRAPPIST-1 from \cite{Agol2021} and assume that these exoplanets are in a 1:1 spin-orbit resonance, i.e, facing the star with the same hemisphere. For TRAPPIST-1\,b and e, we adopt an initial stellar age of 50\,Myr, while for TRAPPIST-1 $\alpha$ we adopt an age of 1\,Gyr. We deviate in the latter case to circumvent the super-luminous pre-main sequence phase of low-mass M-stars for the evolutionary models, but still allow a sufficiently high equilibrium temperature above the silicate solidus for the static models to meaningfully interface with the evolutionary models. 

\setlength{\tabcolsep}{14pt}
\begin{deluxetable*}{llc}
\tabletypesize{\scriptsize}
\tablecaption{Initial parameters for exoplanet model comparison\label{tab:exo_params}}
\tablehead{
\colhead{Parameter} & \colhead{Value} & \colhead{References}
}
\startdata
Starting simulation time [Myr], TRAPPIST 1-b & 50 & -- \\
\phantom{Starting simulation time [Myr],} TRAPPIST 1-e & 50 & -- \\
\phantom{Starting simulation time [Myr],} TRAPPIST 1-$\alpha$ & 1000 & -- \\
Stellar mass [$M_\odot$] & $0.09$ & \citet{Agol2021} \\
Orbital period [days], 1-b & $1.510$ & \citet{Agol2021} \\
\phantom{Orbital period [days],} 1-e & $6.101$ & \citet{Agol2021} \\
\phantom{Orbital period [days],} 1-$\alpha$ & $0.02136$ & fictitious lava world \\
Eccentricity & 0 & --\\ 
Semimajor axis [AU], 1-b & $1.154 \times 10^{-2}$ & \citet{Agol2021} \\
\phantom{Semimajor axis [AU],} 1-e & $2.925 \times 10^{-2}$ & \citet{Agol2021} \\
\phantom{Semimajor axis [AU],} 1-$\alpha$ & $6.75 \times 10^{-4}$ & -- \\
Bond albedo & $0.1$ & \citet{Hamano2015} \\
Planet mass [$M_\oplus$] & $1.0$ & --\\
Planet radius [$R_\oplus$] & $1.0$ & -- \\
Core radius [fraction] & 0.55 & \citet{McDonough1995}\\
T$_{\mathrm{mantle}}^{\mathrm{init}}$ & fully molten & -- \\
Mantle composition & BSE & \citet{vanBuchem2023}\\
Initial oxygen fugacity at 1600 K & IW+4 & --\\
H [kg] / H$_2$O [kg]  & $4.7\times10^{20}$ /  $4.2\times10^{21}$ & --\\
C [kg] / CO$_{2}$ [kg] & $2.73\times10^{20}$ / $1\times10^{21}$ &--\\
\enddata
\tablecomments{Stellar and orbital parameters are adopted from \citet{Agol2021}. The hypothetical planet 1-$\alpha$ uses parameters defined in this work. Volatile inventories and oxidation state assumptions follow \citet{Hamano2015}.}
\label{tab:inputs_trappist1}
\end{deluxetable*}

The stellar input for these exoplanet cases will be represented as a blackbody spectrum whose luminosity, effective temperature, and radius evolve with time according to the stellar parameters and evolution from the \citet{Baraffe2015} evolutionary tracks (provided as a supplemental file on the \chili GitHub repository) to remove the sensitivity of planetary evolution to the choice of stellar model. All simulations should terminate when the mantle solidifies or when they reach the estimated current age of the system (7.6\,Gyr). As in the Solar System cases, each model will retain its natural definition of mantle solidification, be initialized in a fully-molten state, assume a BSE mantle composition, neglect the radiative effects of clouds, and assume a constant Bond albedo of 0.1.

The initial volatile inventories will consist of specified abundances of \ch{CO2} and \ch{H2O} in kilograms (or equivalent H and C inventories). The initial oxygen fugacity (or equivalently Fe$^{3+}$/Fe$_T$) will be assigned an oxidizing value of IW+4, as per the Solar System cases described in Section~\ref{sec:solarsystem}.  Initial volatile abundances will be identical across the three exoplanet cases (Table \ref{tab:exo_params}). 

To enable model intercomparison, we require the same time-evolving outputs as specified for the Earth and Venus models in Section \ref{sec:solarsystem} and listed in Table \ref{tab:out}. The structure and format of the files that should be provided by all evolutionary models is described in Section \ref{sec:output} and repeated in the GitHub online repository, with example output. As for the Solar System cases, comparison between evolutionary and static models will occur at fixed times (Table \ref{tab:snapshots_times}) and following the procedure specified in Section \ref{sec:solarsystem}. 

We hypothesize that TRAPPIST-1\,$\alpha$ is likely to undergo substantial atmospheric loss, leading to an overall increase in mean molecular weight as lighter volatile species are preferentially removed by fractionation \citep{luger2015extreme}. In comparison, TRAPPIST-1\,b is expected to undergo significant cooling accompanied by the depletion of a large fraction of its volatile inventory. TRAPPIST-1\,e is anticipated to cool while retaining a greater portion of its original volatile constituents due to its decreased exposure to ionizing stellar radiation.

\subsection{Static Models} 
\label{sec:staticmodels}

For the static models, outputs of the evolution model at specific times $\tau_i$ will be used as inputs, with the chosen ages summarized in Table \ref{tab:snapshots_times}. We will first use the temporal evolution of the surface temperature of TRAPPIST-1\,e across all evolutionary models and from this ensemble select \emph{three} characteristic ages that describe important evolutionary transitions for a specific planet case. At each of those ages $\tau_i$, we will select the evolutionary models which return the coolest and hottest surface temperature as inputs for the static model comparison. These can be different evolutionary models at different $\tau_i$, as we aim to bracket the possible thermal range at these ages. This will result in 18 sets of parameters for the exoplanet test cases (3 planets $\times$ 3 ages $\times$ 2 evolutionary models) and 12 sets of parameters for the Earth and Venus test cases (2 planets $\times$ 3 ages $\times$ 2 evolutionary models): a total of 30 cases. Static output is not required for the Earth grid (Case 3), only for the nominal Earth and Venus and exoplanet cases (Cases 1 and 2).

The static models will take the inputs  and outputs described in Table \ref{tab:static_params}, where possible, to compare with time-evolution models.  Based on their capabilities, static models are categorized as: (i) magma ocean surface chemistry models, which do not capture a physical layer of a magma ocean planet, but rather report the chemical speciation at the interface between the atmosphere and the magma ocean; or (ii) atmospheric structure models, where the vertical structure of the atmosphere is resolved. Since the capabilities and emphases vary across static models, some flexibility is needed in terms of their inputs to both demonstrate their capabilities and best match the outputs from the evolutionary models.

For the magma ocean surface models (i), we expect these models to, at least, take $p_{\ch{H2O}}$ as inputs from evolutionary models, or ideally the bulk elemental abundances in the atmosphere. They should at least return the partial pressures of individual species. 

For the atmospheric structure models (ii) with self-consistent radiative-convective balance, the required inputs in Table \ref{tab:static_params} should suffice. If not, then additional outputs such as $T_{\rm surf}$ (Table \ref{tab:out}) can be used. In terms of the chemical composition for the static atmospheric structure models, they should, in principle, take at least the partial pressure of \ch{H2O} as input from an evolution model, or more species $x$ if offered by the evolution model and allowed by the capabilities of the static model. If an atmospheric structure model also calculates atmospheric speciation, then the elemental abundances are also acceptable inputs. If the composition is incomplete (i.e. if $\sum p_x<p_{\rm surf}$), then the input partial pressures should be normalized via
\begin{equation}
    p_{x, \rm norm} = p_{x}\cdot \frac{p_{\rm surf}}{\sum p_{x}}.
    \label{eq:norm}
\end{equation}
These static structure models should output the pressure, temperature, and composition as a function of distance $z$ from the magma ocean surface.

\rev{Static models are not expected to model interior properties such as viscosity, melt fraction, or radiogenic heating; these quantities are relevant only to evolutionary models. However, static models may ingest as many input parameters from the evolutionary model output as their framework supports, and should state which input quantities they make use of in the accompanying notes file.} This framework will enable systematic comparison between time-dependent and equilibrium states, and provide a foundation for quantifying how stellar evolution and atmospheric escape jointly shape the thermal and compositional evolution of rocky planets. Outputs from the static models at times defined by Table~\ref{tab:snapshots_times} will enable consolidation of all simulation results, and a homogeneous intercomparison between evolutionary and static calculations. 
 
\setlength{\tabcolsep}{14pt}

\begin{deluxetable*}{lcccccc}

\tabletypesize{\scriptsize}
\tablecaption{Characteristic evolutionary ages. Snapshots taken at a subset of these times will generate input parameters for the static models. Not all of these times will necessarily be important in all planet cases; they serve as reference points for the interface between the evolutionary and static models. Evolutionary models should store output variables at times close to these ages where possible.}
\label{tab:snapshots_times}
\tablehead{
 \colhead{$\tau_3$} & \colhead{$\tau_4$} & \colhead{$\tau_5$} & \colhead{$\tau_6$} & \colhead{$\tau_7$} & \colhead{$\tau_8$} & \colhead{$\tau_9$}
}
\startdata
     $10^{3}$ yrs & $10^{4}$ yrs & $10^{5}$ yrs& $10^{6}$ yrs& $10^{7}$ yrs& $10^{8}$ yrs & $10^{9}$ yrs\\
\enddata
\end{deluxetable*}

\rev{The characteristic ages $\tau_3$ through $\tau_9$ span the range of expected magma ocean lifetimes across the parameter space explored by \chili. The three specific ages selected as inputs for the static models will be chosen post-hoc from the evolutionary model ensemble, targeting times that capture key transitions in the solidification sequence (e.g., onset of rapid crystallization, rheological transition, and near-complete solidification).}

\begin{table*}
\centering
\begin{tabular}{c|c|c}
\tabletypesize{\scriptsize}
&
\begin{minipage}[t]{5cm}
    \textbf{(i) Magma ocean surface chemistry models}
\end{minipage}
 & 
 \begin{minipage}[t]{5cm}
    \textbf{(ii) Atmospheric structure models}
\end{minipage}
  \\ \hline
\begin{minipage}[t]{3cm}
    Required inputs from evolutionary models 
\end{minipage}
&

\begin{minipage}[t]{5cm}
$p_{\rm surf}$, $T_{\rm surf}$
\\
At least $p_{H_2O}$, ideally bulk atmospheric C/H/O ratios. 
\end{minipage} &

\begin{minipage}[t]{5cm}
$F_{\rm surf}$, $F_{\rm ASR}$, $p_{\rm surf}$
\\
At least *$p_{x,\rm norm}$, ideally bulk atmospheric C/H/O ratios.  
\end{minipage}\\

&&\\ \hline
\begin{minipage}[t]{3cm}
Required outputs from static models

\end{minipage}
 & 
 
\begin{minipage}[t]{5cm}
$p_x$
\end{minipage} &

\begin{minipage}[t]{5cm}
$p(z), p_x(z), T(z)$
\end{minipage}\\
\end{tabular}
\caption{Required inputs and outputs for static magma ocean chemistry and atmosphere models. *Normalized atmospheric partial pressures of individual species in bars (Equation \ref{eq:norm}).}
\label{tab:static_params}
\end{table*}

\subsection{Preliminary model comparison}
\label{sec:preliminary_results}

The first version of the \chili protocol was developed during an in-person workshop in October 2025. Simultaneously, four evolutionary codes were used to simulate preliminary cases for Earth and TRAPPIST-1\,b: GOOEY, PROTEUS, PACMAN and CAMO. Additionally, two static  models (MOAChi and LavaAtmos 2.0) were compared to the evolutionary models for TRAPPIST-1\,b at $\tau_5=10^5$ yrs. The outputs from the PROTEUS TRAPPIST-1\,b run were used as inputs for the static models. \rev{PROTEUS was chosen because it provided the most complete set of required output variables at the time of the workshop. Only a single characteristic age ($\tau_5 = 10^5$\,yr) was used for the static model inputs in this preliminary demonstration due to workshop time constraints; the full intercomparison will employ multiple $\tau$ values as specified in Table~\ref{tab:snapshots_times}.} PROTEUS predicts a molten surface of $T_{\rm surf}\sim1800$\,K for TRAPPIST-1\,b at $\tau_5=10^5$\,yrs. 

Results of the four Earth evolutionary models are shown in Figure \ref{fig:earth_evolve}. Results of all six TRAPPIST-1\,b models (evolutionary + static) are shown in Figure \ref{fig:T1b_evolve}. A goal of this preliminary test is to ensure that the initial conditions specified in this protocol are attainable by each model, and to ensure that they produce reasonable results from a non-exhaustive selection of distinct models. The various assumptions made for the preliminary model runs are described in the following subsections. 

The following description is meant to illustrate the comparative power of the model comparison and provide some initial tests of how the models may differ, not replace the intercomparison itself. Therefore, we provide only brief descriptions and speculative discussions on the possible origins of the model differences. The full intercomparison will serve to do this exercise in detail.

\subsubsection{Model specifics: GOOEY, PROTEUS, PACMAN, CAMO}

Each evolutionary model is free to adopt a different critical melt fraction that governs mantle viscosity behavior. The GOOEY model assumes a critical melt fraction of 40\%. GOOEY includes only \ch{H2O} and O$_2$ gas evolution, unlike the other models that include carbon-species and other hydrogen-bearing species. Oxygen fugacity evolves during the evolution, with an initial value prescribed by the Fe$^{+3}$/Fe$_{T}$ abundance, which is taken here to be 0.05 for Earth and 0.17 for TRAPPIST-1\,b. GOOEY also accounts for atmospheric loss through its entire evolution using energy-limited escape, with fractionation between H and O allowed and an assumed efficiency of 30\%. We adopt an initial mantle temperature of 3600\,K. 

In PROTEUS, we set the critical melt fraction (which defines the rheological transition) at 50 wt\% melt fraction. Volatile species included were \ch{H2O}, \ch{CO2}, \ch{H2}, \ch{CH4}, \ch{O2}, and \ch{CO}. Escape is included with a 10\% energy efficiency; the hydrodynamic outflow composition is equal to the bulk atmosphere (i.e., not the bulk planet volatile inventory), and we make corrections for Roche lobe overflow \citep{Erkaev_Rochelobe_2007}. PROTEUS adopts the \citet{spada_radius_2013} stellar evolution tracks for these tests, and with these finds Earth's 50\,Myr bolometric instellation to be $\rm 1010\,W\,m^{-2}$, which is $+9.7\%$ larger than the $\rm 920\,W\,m^{-2}$ predicted by models which adopt the \citet{Baraffe2015} stellar tracks. Oxygen fugacity in the mantle is fixed relative to iron-w\"ustite, at IW+4 throughout the evolution. PROTEUS adopts an isochemical atmospheric profile in this test.

In the PACMAN model, we parameterize mantle solidification by analytically solving for the intercept between the solidus and mantle adiabat as described in \citet{krissansen-totton2021}. The model version used in this intercomparison, PACMAN-P, includes atmospheric volatile species H$_2$O, CO$_2$, O$_2$, CO, CH$_4$, and H$_2$ \citep{Krissansen-Totton_2024}. Fractionated escape was allowed for H-, C-, and O-bearing species following the procedure detailed in \citet{Krissansen-Totton_2024}. This model also calculates self-consistent redox evolution, therefore initial free oxygen inventory was adjusted to produce a final mantle redox state near the IW+4 buffer, consistent with Earth's modern mantle.  

In the CAMO model, the critical melt fraction is fixed at 40$\%$, by temperature, marking the transition between liquid-like and solid-like mantle behavior. The atmosphere is composed of H$_2$O, CO$_2$, and a constant background of N$_2$. Atmospheric escape is calculated from an energy-limited escape flux of water following the parameterization of \citet{Hamano2015}. The mantle solidus is based on the compositional dependence described by \citet{Andrault2011}, while the viscosity follows the temperature and melt-dependent parametrization by \citet{Salvador2017} and references therein. In this test, the model was run with isochemical vertical atmospheric profiles.
\rev{Plots of $T_{\rm surf}$ versus $\phi$ for all models are planned for the full intercomparison analysis.}

It was necessary for some of the users to convert the input conditions into different quantities relevant to their own code requirements, and to do some post-processing of results in order to provide equivalent quantities. For instance, outputs for melt fraction ($\phi$) and mantle potential temperature (T$_{p}$) are defined differently between the four evolutionary models. The PROTEUS model, for instance, had to modify their methodology for computing ($\phi$), which that code defines in terms of mass fraction, because the protocol requests melt fraction by volume. 

One issue that highlights the need for the \chili project is that models differently define their endpoint of the magma ocean stage. Both GOOEY and PACMAN define the end of the magma ocean as the time when the surface temperature reaches the solidus temperature. However, their respective solidus temperatures disagree (GOOEY: 1420\,K, PACMAN: 1473\,K). PROTEUS and CAMO respectively define the endpoint of the magma ocean by melt fraction thresholds of 2\% (by mass) and 0\% (by volume). 

\subsubsection{Model specifics: MOAChi, LavAtmos\,2.0}

MOAChi is a coupled atmosphere-interior structure model. For the example test cases in this protocol, light modification was done to better match MOAChi inputs with the outputs from the evolutionary models. Instead of receiving the bulk C and H abundances as inputs, MOAChi takes the atmospheric elemental abundances and total surface pressure $p_{\rm surf}$, as well as $F_{\rm surf}$ and $F_{\rm ASR}$. This allows for a more direct comparison of modeled atmospheric structure between PROTEUS and MOAChi.

LavAtmos\,2.0 is a magma ocean surface chemistry model, outputting the rock vapor and volatile composition at the magma ocean-atmosphere interface. LavAtmos requires a phosphorus abundance. We set this to a value of $10^{-20}$ to mimic a phosphorus-poor scenario. Both the input and output parameters required for each of the static models are listed in Table \ref{tab:static_params}. With LavAtmos, we test how the partial pressures of outgassed species behave in a temperature range around this value. LavAtmos takes the presence of a volatile envelope from only C-H-N-P-S into account, but cannot account for solubility of species in the magma. Therefore all the oxygen in the model originates from the oxygen in the BSE melt and solubility of species is not accounted for. This means that adding a pre-existing atmosphere of \ch{H2O} and \ch{CO2} as obtained in the evolution models is not possible. If we assume that our volatile envelope is composed of carbon and oxygen only, this leads to an atmosphere primarily composed CO. Hence, we ran a test case for a pure hydrogen envelope. For this test case we scaled the atmospheric pressure to the hydrogen partial pressure from the PROTEUS output at $\tau_5=10^5$\,yrs, corresponding to a 107\,bar surface pressure. This leads to 0.989\,bar of \ch{H2O} predicted in the atmosphere with LavAtmos.

\subsubsection{Preliminary results}

\begin{figure*}[tbh!]
    \centering
    \includegraphics[width=\linewidth]{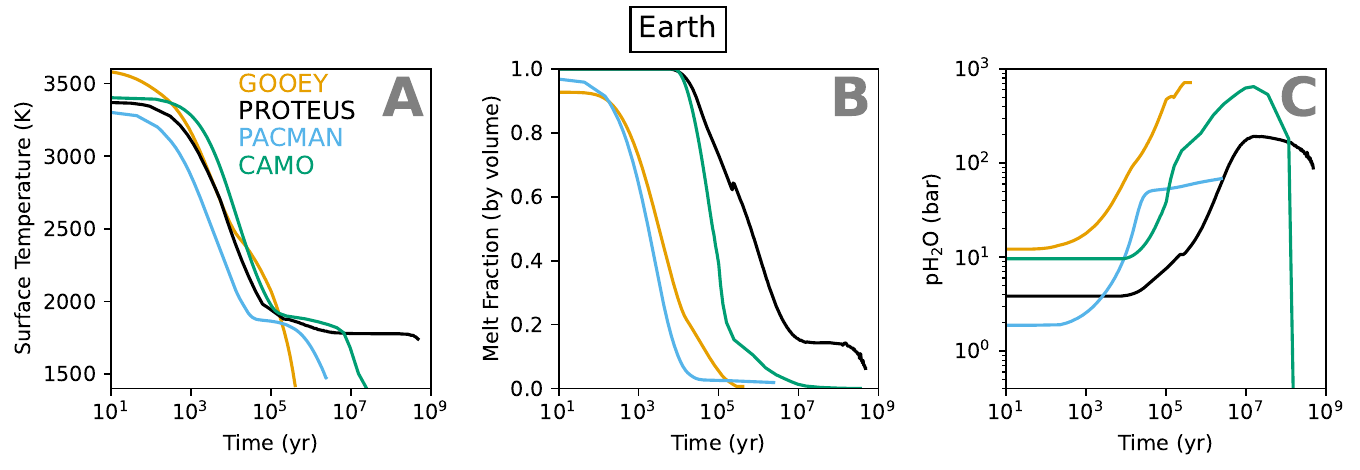}
    \caption{Evolutionary models example test outputs for the nominal Earth case, with inputs defined in Table \ref{tab:inputs_solar_system}. The figure shows surface temperature in Kelvin (panel A, left), volumetric melt fraction of the whole mantle \rev{(dimensionless volume fraction)} (panel B, center), and partial pressure of \ch{H2O} in bars (panel C, right). All model results are truncated at the model-specific endpoint time. Only a subset of the evolutionary models that will participate in \chili are shown here.}
    \label{fig:earth_evolve}
\end{figure*}

For the Earth case (Fig.~\ref{fig:earth_evolve}), we show preliminary results for three parameters that we will use to compare models: the surface temperature, the melt fraction, and the partial pressure of atmospheric H$_2$O. All models start at a temperature necessary to produce a fully molten mantle. Although the initial surface temperature of GOOEY is higher than the other models, it reaches the end of the magma ocean state first. The surface temperature evolution of PROTEUS and PACMAN are initially very similar, but PROTEUS reaches the end of the magma ocean at a higher surface temperature and the models produce mantle melt fractions that differ significantly. GOOEY and PACMAN produce similar melt fractions, although the surface temperature of GOOEY is consistently hotter than PACMAN. CAMO has an initial surface temperature similar to PROTEUS and solidifies at a similar time as PACMAN. GOOEY, PACMAN, and CAMO solidify on timescales of $\sim 10^5$ to $10^6$ yrs. CAMO deviates by showing a steep decrease of melt fraction at about $10^5$ yr. PROTEUS deviates the strongest from the other three models in both surface temperature and melt fraction evolution; its simulation starts to solidify earlier than CAMO, but then solidification progresses slowly, reaching a plateau of melt fraction of $\sim$0.15 vol\% at $\sim10^7$\,yr during which radiative equilibrium is attained, and solidification is only reached after $\sim10^8$\,yr. GOOEY and CAMO produce similar \ch{H2O} partial pressures over time, reaching $\sim700$\,bar. PACMAN and PROTEUS instead produce on the order of $\sim100$\,bar of \ch{H2O} in the atmosphere upon solidification. The time evolution for all models correlate with their melt fraction evolution, with \ch{H2O} being outgassed as the mantle solidifies. PROTEUS deviates accordingly, showing some decrease in pressure as the simulated timescales are sufficiently long for atmospheric escape to play an important role.
\rev{These differences are likely related to distinct physical assumptions in each model. GOOEY achieves faster solidification partly because its gray atmosphere treatment may permit more efficient radiative cooling. PROTEUS and CAMO exhibit extended solidification timescales, possibly because their non-gray radiative transfer models capture deep radiative layers that insulate the interior and slow cooling. However, CAMO uses a 0-D interior model, whereas PROTEUS employs a 1-D interior structure, which may also contribute to differences in their evolution outcome. PACMAN produces intermediate solidification times, likely reflecting its self-consistent redox evolution and fractionating atmospheric escape that jointly modulate the atmospheric opacity and cooling rate. A detailed attribution of these differences will be the subject of the intercomparison results papers; the protocol presented here serves to illustrate that significant model-to-model differences exist and motivate systematic comparison.}

\begin{figure*}[tbh!]
    \centering
    \includegraphics[width=1\linewidth]{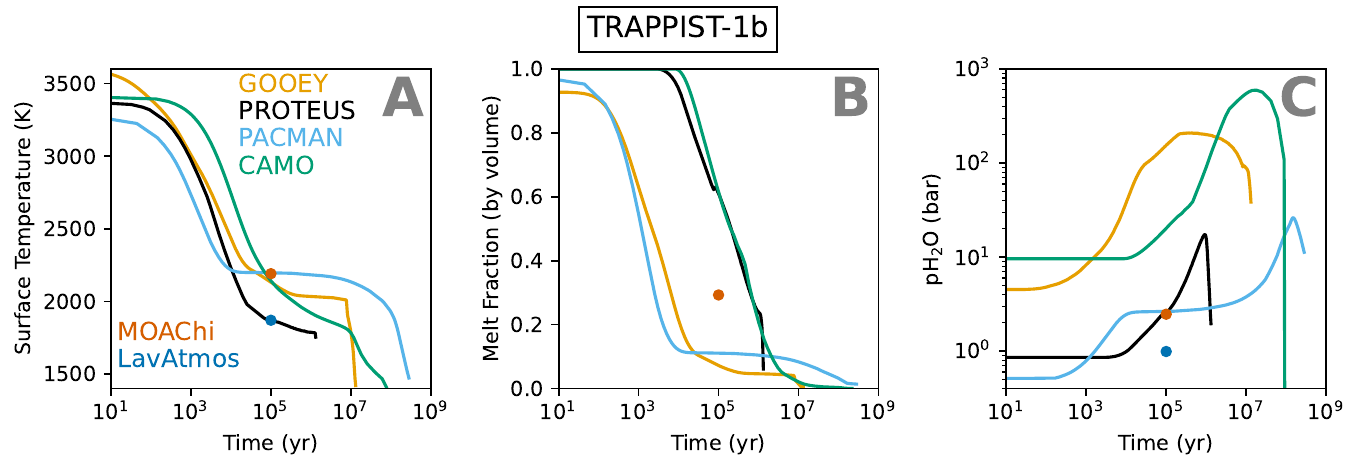}
    \caption{Evolutionary models example test outputs for the TRAPPIST-1\,b case, with inputs defined in Table \ref{tab:inputs_trappist1}. The figure shows surface temperature in Kelvin (panel A, left), volumetric melt fraction of the whole mantle \rev{(dimensionless volume fraction)} (panel B, center), and partial pressure of \ch{H2O} in bars (panel C, right). Dots show the predictions from two static models, MOAChi and LavAtmos at $10^5$\,yrs.}
    \label{fig:T1b_evolve}
\end{figure*}

For the TRAPPIST-1\,b exoplanet case, we show preliminary results for the same three output variables as in the Earth case to compare the models (Figure \ref{fig:T1b_evolve}).  We also show predictions from the static MOACHi and LavAtmos models as dots in Fig. \ref{fig:T1b_evolve}; these two models only make predictions at a single time, set to $\tau_5=10^5$\,yr. We also adopted the PROTEUS evolutionary output as the input for these static model calculations. While MOAChi is able to provide surface temperature, melt fraction and water partial pressure as outputs, LavAtmos computes the atmospheric chemical composition at a given temperature: therefore, we report only the temperature in panel A and the water partial pressure for this model in panel C of Fig. \ref{fig:T1b_evolve}.

In Figure \ref{fig:T1b_evolve}A, we show the time evolution of surface temperature calculated by GOOEY, PROTEUS, PACMAN, and CAMO. The initial temperatures differ due to the varying initial environments for each model, despite the consistency with the standardized initial inputs. Again, GOOEY and CAMO begin at higher surface temperatures, whereas PACMAN and PROTEUS initially show lower surface temperatures. Final surface temperatures vary with respect to the differences in how each model defines the end of the magma ocean phase; PROTEUS and CAMO allow higher temperatures at the solid surface, while GOOEY and PACMAN end with lower surface temperatures. The general trend of surface temperature evolution is similar between models, indicating adherence to the same underlying physics. Additionally, differences between when the mantle solidifies for Earth and TRAPPIST-1\,b hints at these models' sensitivity to differing stellar spectra, with GOOEY and PACMAN indicating longer magma oceans for TRAPPIST-1\,b, and CAMO and PROTEUS showing a significantly quicker magma ocean solidification compared to the Earth test case. This is likely due to their treatment of a radiative layer in the deep interior.

In Figure \ref{fig:T1b_evolve}B, the time evolution of the volumetric melt fraction is plotted for each model, illustrating the differences in how each model treats solidification. GOOEY and PACMAN show earlier rapid solidification followed by later slower solidification below a melt fraction of $\sim$ 0.1-0.2, respectively. In contrast, PROTEUS and CAMO indicate that the mantle remains fully molten for $\sim10^{5}$\,yrs, before then rapidly solidifying. 

Figure \ref{fig:T1b_evolve}C shows the time evolution of the partial pressure of atmospheric H$_2$O. It is important to note that PROTEUS and PACMAN allow various hydrogen-containing atmospheric species, while GOOEY and CAMO allow only H$_2$O and therefore report higher $p$H$_2$O early in the evolution. The models treat dissolution of H$_2$O into mantle and atmospheric escape in varying ways, which contribute to the different outcomes between models. GOOEY and PACMAN allow fractionating escape, while PROTEUS and CAMO adopt non-fractionating escape (Figure \ref{fig:overview_models}; Section \ref{sec:models}). The models all show similar behavior of increasing $p$H$_2$O as melt fraction decreases, as H$_2$O is outgassed from the mantle, followed by decreasing $p$H$_2$O as atmospheric escape depletes the atmosphere over a longer time-scale. $p$H$_2$O at the model endpoints differ by two orders of magnitude, from $\sim300$\,bar to $\sim20$\,bar.

The red and blue dots in Fig.~\ref{fig:T1b_evolve} illustrate calculations from two static models. MOAChi used PROTEUS inputs, however, its predicted surface temperature is higher than both PROTEUS and CAMO (Fig.~\ref{fig:T1b_evolve}A), but approximately similar to GOOEY and PACMAN. Despite a hotter interior than PROTEUS, MOAChi estimates a shallower magma ocean at $\sim$30 vol\% (Fig.~\ref{fig:T1b_evolve}B), higher than GOOEY, PACMAN and CAMO, but lower than PROTEUS. The water partial pressure $p$H$_2$O is consistent between MOAChi and PROTEUS, which indicates atmospheric speciation operates similarly between the two models. LavaAtmos predicts a lower water vapour partial pressure, which is explained by the workings of the model and the chosen setup, which uses an H$_2$ dominated atmosphere, without dissolution.

\begin{figure*}[bth!]
    \centering
    \includegraphics[width=0.65\linewidth]{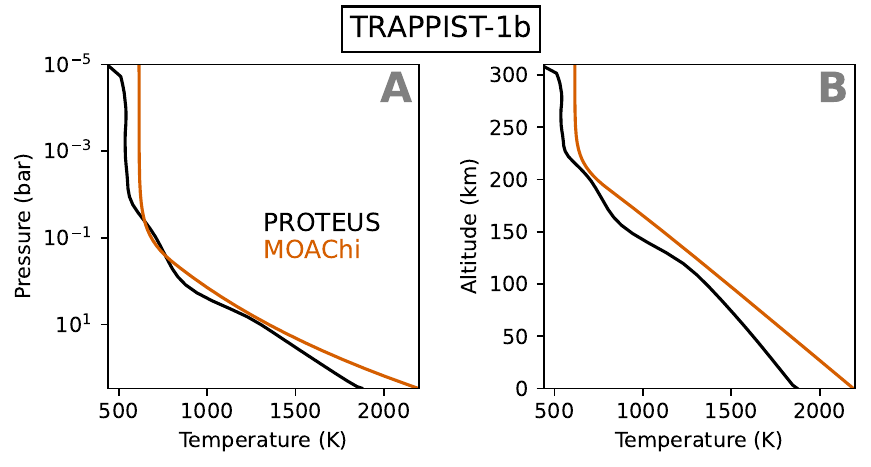}
    \caption{Static model example test with MOAChi for the TRAPPIST-1\,b case at $\tau_5 = 10^5$ yr for the output of the evolutionary model PROTEUS. The figure shows surface temperature as a function of pressure (panel A) and altitude (panel B).} 
    \label{fig:T1b_static}
\end{figure*}

Finally, Fig. \ref{fig:T1b_static} compares the predicted atmospheric structure calculated by MOAChi and PROTEUS at $\tau_5=10^5$\,yr. Both profiles qualitatively agree with each other, but show some differences near the surface. A strong internal heat flux keeps both atmospheric profiles close to adiabatic, although MOAChi predicts a steeper temperature gradient and hotter lower atmosphere. Using a non-grey radiative transfer scheme likely accounts for the sub-structures in the PROTEUS atmosphere that MOAChi does not predict, yielding a somewhat shallower lapse rate. Finally, the MOAChi atmosphere is also somewhat more extended, which is consistent with its hotter, puffier lower atmosphere.

In summary, comparison between the Earth and TRAPPIST-1b test cases illustrates the order-of-magnitude deviation in model predictions, with comparatively reversed differences between Earth and TRAPPIST-1\,b for PROTEUS, for example. Overall the trends are physically justifiable: the planets cool, outgas volatiles, which can escape to space. However, the varying timescales between models indicate substantial variation in how the underlying physics and chemistry are treated. The initial test for the atmospheric structure reveal subtle but meaningful differences in the model outputs, which can affect atmospheric observables for exoplanet observations. Our results therefore illustrate the need for a detailed intercomparison project, which \chili will perform. We will discern in detail which treatments of the underlying physics and chemistry affect the model outcomes within the planned intercomparison papers.

\section{Output diagnostics and repositories}
\label{sec:output}

The data and scripts resulting from the intercomparison and the analysis files will be hosted on the CHILI GitHub repository\footnote{\url{https://github.com/projectcuisines/chili}}.  After conclusion of the intercomparison and publication of the results, all data and files will also be archived at the permanent CUISINES repository website\footnote{\url{https://ckan.emac.gsfc.nasa.gov/organization/cuisines-chili}}. This includes the input and output files from both the evolutionary and static models. Because each model takes in a wide range of parameters in different ways, we do not have a specific input file that all models will take, but rather each model implements the input parameters from tables \ref{tab:inputs_solar_system} and \ref{tab:exo_params}, respectively, for the Solar System and exoplanet cases. For the static models, we will have an internal output file following Table \ref{tab:snapshots_times} with crucial parameters from the evolutionary models at specific times. This will include melt fraction, atmospheric pressure and surface temperature. We encourage the reader to visit the \chili GitHub repository to obtain all necessary information for participation in the intercomparison, example files, formatting guidelines, and timelines for participation. The GitHub repository will be a central hub for the intercomparison and will be continuously updated until the completion of the project. All repository versions, including that corresponding to the protocol paper, are permanently archived on the \rev{\href{https://doi.org/10.5281/zenodo.17312304}{CHILI Zenodo Archive}} \citep{chili_zenodo}. The \chili project is open to the community and we encourage the participation of all models which fall within the above-defined descriptions of magma ocean models.

\subsection{Output Files}

Many atmospheric and interior quantities may be important to understand and verify differences in the underlying physical and chemical principles that are modeled by the different codes. The example figures from the protocol manuscript, the scripts to produce them, and the underlying data from the protocol can be found in the \href{https://github.com/projectcuisines/chili}{CHILI GitHub repository}. Please note that the output files for the actual intercomparison will be more expansive than for the protocol, to give us room to evaluate the differences between the models. If a model does not output a specific parameter, the output file should still include the respective column, but \rev{filled with NaN}.

The results in CSV format, along with the model inputs and scripts used to analyze and visualize the data will be made available on the \href{https://github.com/projectcuisines/chili}{\chili GitHub repository}. Example cases can be found in sub-folders, illustrating how each model should report its results. Each participating model should deposit data in a sub-folder of the repository. The data output CSV file(s) should be accompanied by a text note file that itemizes all noteworthy characteristics of a code that deviate from published code descriptions. For example, changes to the code that were made to adhere to the CHILI protocol, as-of-yet unpublished updates to a code, or conversions of output data to comply with the required units. In addition, all code folders should contain the exact configuration files used to generate the corresponding simulation output data, and describe in the TXT file how the code may be obtained, stating the code version adopted, and a contact method for one of the participating authors. Optimally, codes should be made available through a Zenodo archive and on an open-source code repository. 

The optimal way to submit data is by opening a new branch to the repository with name formatted as \texttt{modelname-data}, modifying the branch accordingly, and then submitting a \textit{Pull Request} to the \textsc{main} branch with a description of the changes. This will enable adherence to the data guidelines and  participation of a large number of diverse community codes. All data must comprise $<10$\,MB to not overburden the repository. The GitHub repository also lists contact information for the \chili organizer team, who can clarify questions and extend invitations to the project mailing lists. 

\rev{The detailed output file format specifications, column headers, and file naming conventions for both evolutionary and static models are provided in Appendix~\ref{sec:appendix_output}.}

\section{Potential impact}
CHILI will provide the first systematic intercomparison of magma ocean evolution models for Solar System bodies and exoplanets. We anticipate the following points of scientific impact:
\begin{itemize}
    \item Determining the spread in estimates for Earth's magma ocean solidification time, excluding the influence of tidal forcing. Investigating the reasons behind this spread (or lack thereof) may provide insights on Hadean Earth system evolution and the emergence of conditions suitable for the origin of life. 
    \item Investigating the combinations of initial C and H inventories that result in potentially habitable conditions on Earth. The extent to which models agree on this question may inform our understanding of plausible initial volatile inventories for the Earth, hence linking planetary evolution to formation theories.
    \item Understanding the Earth/Venus dichotomy. The extent to which magma ocean evolution models agree on Venus's solidification time and the possibility of ocean formation may inform understanding of Venus's early evolution, and when Venus's climate evolution diverged from that of the Earth.
    \item The spread in magma ocean solidification times and volatile evolution histories for TRAPPIST-1\,b and e will provide insights on terrestrial planet evolution around late M-dwarfs, in particular on whether they can hold onto secondary atmospheres. 
    \item The ensemble of model outcomes may be used to assess the likelihood of abiotic oxygen build-up/removal during the stellar pre-main sequence phase, and corresponding prospects for the emergence of habitable conditions.
    \item The comparative simulations of a representative ultra-hot Earth-sized exoplanet (TRAPPIST-1\,$\alpha$) will provide insights for the ongoing JWST characterizations of hot ``lava worlds", including distinguishing between rock vapor and volatile-rich atmospheres, and the extent of mantle melting under extreme instellation. 
\end{itemize}

We plan for the intercomparison to be undertaken with a three-branch approach: (i) evolutionary study of Solar System planets, (ii) evolutionary study of exoplanets, and (iii) static modeling study of both Solar System planets and exoplanets. Through these three branches, we expect to test and sample all atmosphere-interior models as listed in Fig. \ref{fig:overview_models}. 

\section{Summary} \label{sec:summary}

In this manuscript we presented the Version 1.0 of the Coupled atmospHere Interior modeL Intercomparison (CHILI) project, which will aim to compare multiple codes used and developed in the community to model the evolution and atmospheres of the magma ocean phase on Solar System planets and on exoplanets. We have detailed the main motivation and technical aspects of the intercomparison, and outlined the required input and output data, and anticipated impact. We presented preliminary example results from the intercomparison tests, which highlight noticeable differences among the models, which justify the rigor of our experimental protocol to break down the causes of these differences. To facilitate wide community engagement and easy reproducibility, the full results will be presented in separate follow-up works, as is commonly done within the CUISINES framework \citep{Sohl24_cuisines}. We encourage interested colleagues to join the intercomparison by reaching out to the CHILI organizers and submitting their code details and output data through the \href{https://github.com/projectcuisines/chili}{CHILI GitHub repository}.

\onecolumngrid
\pagebreak

\begin{acknowledgments}
\nolinenumbers
CRediT author statements: Conceptualization: TL, LS, JKT, YM, DS, LSo.
Methodology: TL, LS, JKT, YM, JC, HN, EP, MS, HP, PB, LN, GN, AS, BP.
Software: TL, LS, JKT, YM, HN, EP, MS, HP, JL, PB, LN, LJ, GN, AS, AP, BP.
Validation: TL, LS, JKT, YM.
Formal analysis: HN, AZ, AP, BP.
Investigation:  LJ, LS, AP.
Resources: TL, LS, JKT, YM, DS
Data Curation: TL, LS, JKT, AZ.
Writing – Original Draft: TL, LS, JKT, YM, DS, JC, HN, EP, MS, HP , LJ, JP, PB, LN, AZ, AS, AP, BP, YM.
Writing – Review \& Editing: TL, LS, JKT, YM, DS, TJF, JP, HN, AZ, AS, BP, HP, JL, MM.
Visualization: TL, AZ.
Supervision: TL, LS, JKT, YM, DS, TJF, JL.
Project Administration: TL, DS.
Funding Acquisition: TL, TJF.
\chili belongs to the CUISINES framework, a Nexus for Exoplanet System Science (NExSS) science working group. Authors acknowledge support from the GSFC Sellers Exoplanet Environments Collaboration (SEEC), which is funded in part by the NASA Planetary Science Divisions Internal Scientist Funding Model (ISFM). We thank the Netherlands eScience Center (PROTEUS project, NLESC.\-OEC.\-2023.\-017) and the Lorentz Center for funding and support in the organisation of the workshop `Atmospheric and interior evolution of planetary magma oceans' in October 2025 in Leiden, during which the majority of this protocol manuscript was compiled. TL was supported by the Branco Weiss Foundation, the Alfred P. Sloan Foundation (AEThER project, G-2025-25284), NASA’s Nexus for Exoplanet System Science research coordination network (Alien Earths project, 80NSSC21K0593), the NWO NWA-ORC PRELIFE Consortium (NWA.1630.23.013), and the European Research Council (ERC) under the European Union's Horizon Europe research and innovation programme (101219807, MagmaWorlds). LN and PB are funded by the European Union (ERC, DIVERSE, 101087755). Views and opinions expressed are however those of the author(s) only and do not necessarily reflect those of the European Union or the European Research Council Executive Agency. Neither the European Union nor the granting authority can be held responsible for them. JKT and AP were supported by a NASA Astrophysics Decadal Survey Precursor Science grant 80NSSC23K1471, the Virtual Planetary Laboratory, a member of the NASA Nexus for Exoplanet System Science (NExSS), funded via the NASA Astrobiology Program grant No. 80NSSC23K1398, and the Alfred P. Sloan Foundation under grant No. 2025-25204. JC was supported by NSF CAREER award AST-1847120, NASA Virtual Planetary Laboratory award UWSC10439 and NSF award AGS-2210757. TJF was supported by the SEEC ISFM award 2024 "The CUISINES menu for FY25 and FY26". HN acknowledges support from STFC grant UKRI1184.  HN thanks the Center for Information Technology of the University of Groningen for their support and for providing access to the H\'abr\'ok high performance computing cluster. K.H. was supported by MEXT KAKENHI Grant Number JP22H05150. YM and LJ acknowledge support from the European Research Council (ERC) under the European Union’s Horizon 2020 research and innovation programme (grant agreement no. 101088557, N-GINE). HP and JL acknowledge the support of the French Agence Nationale de la Recherche (ANR), under grant ANR-20- CE49-0009 (project SOUND), the Programme National de Planétologie (PNP), and CNES. 
\end{acknowledgments}


\appendix
\section{Output File Format Specifications} \label{sec:appendix_output}

\rev{This appendix provides the detailed output file format specifications for participating models. The \chili GitHub repository serves as the authoritative and up-to-date reference for these specifications.}

\subsection{\rev{Repository structure}}

\rev{Within the intercomparison directory of the GitHub repository, there is a single \texttt{inputs/} directory and a single \texttt{outputs/} directory. Each model has its own subdirectory within these (e.g., \texttt{inputs/<modelname>/}, \texttt{outputs/<modelname>/}). All code names are written in lowercase letters in filenames, irrespective of their particular capitalization (e.g., \texttt{moachi} instead of \texttt{MOAChi}).}

\rev{Each individual output file should not exceed a file size of 1\,MB, and the total output size for each code cannot exceed 10\,MB.}

\subsection{Evolution model outputs}
Output CSV files for evolutionary models should contain columns with the following quantities, in line with Table \ref{tab:out}. Here, we describe the format of the column headers.

\onecolumngrid
\begin{verbatim}
 t(yr)             Time in years
 T_surf(K)         Surface temperature
 T_pot(K)          Potential temperature
 flux_surf(W/m2)   Net geothermal heat flux from interior to atmosphere
 flux_OLR(W/m2)    Top of atmosphere outgoing longwave radiation
 flux_ASR(W/m2)    Top of atmosphere average absorbed stellar radiation
 phi(vol_frac)     Mantle total volume fraction of melt
 fO2_solid(bar)    Oxygen fugacity of solid mantle
 fO2_melt(bar)     Oxygen fugacity of melt
 thick_surf_bl(m)  Thickness of surface viscous boundary layer
 massC_solid(kg)   Mass of carbon in the solid mantle
 massC_melt(kg)    Mass of carbon in the melt
 massC_atm(kg)     Mass of carbon in the atmosphere
 massH_solid(kg)   Mass of hydrogen in the solid mantle
 massH_melt(kg)    Mass of hydrogen in the melt
 massH_atm(kg)     Mass of hydrogen in the atmosphere
 massO_atm(kg)     Mass of oxygen in the atmosphere
 p_surf(bar)       Total atmospheric surface pressure
 p_H2O(bar)        Partial atmospheric pressure of H2O
 p_CO2(bar)        Partial atmospheric pressure of CO2
 p_CO(bar)         Partial atmospheric pressure of CO
 p_H2(bar)         Partial atmospheric pressure of H2
 p_CH4(bar)        Partial atmospheric pressure of CH4
 p_O2(bar)         Partial atmospheric pressure of O2
 mmw(kg/mol)       Mean molecular weight of the atmosphere
 R_trans(m)        Transit radius of the planet
 R_solid(m)        Radius of the rheological transition in the mantle
 viscosity(Pa.s)   Characteristic viscosity of the mantle
\end{verbatim}

Reported times can be model-specific, but should be sampled finely enough to resolve potentially-rapid solidification, and feature times close to the ages $\tau_i$ in Table \ref{tab:snapshots_times}. \rev{We recommend logarithmically-spaced output with at least 50 points per decade in time, and output times within 1\% of each $\tau_i$ age.} Submitted output files for the evolutionary models should be:
\begin{itemize}
    \item \texttt{evolution-[modelname]-[planet]-data.csv}
    \item \texttt{evolution-[modelname]-earth-grid-H[low,mid,high]-C[low,mid,high]-data.csv}
    \begin{itemize}
        \item  For example: \texttt{evolution-gooey-earth-grid-Hlow-Cmid-data.csv} for the output data of the evolution code GOOEY with H = 1.6 $\times 10^{20}$ kg and C = 2.72 $\times 10^{20}$ kg (Table~\ref{tab:inputs_solar_system}).
    \end{itemize}
    \item \texttt{evolution-[modelname]-notes.txt}
    \item Any other configuration files necessary to reproduce the output data.
\end{itemize}
Some evolutionary models can also produce output variables similar to the static models; e.g. atmospheric profiles. In these cases, the evolutionary models should store these files \rev{with a \texttt{-tau[X]} marker before the \texttt{-data.csv} suffix (e.g., \texttt{evolution-[modelname]-[planet]-tau5-data.csv})}, following the nomenclature of the static models described below, substituting \texttt{[hot,cold]} with the model name.

\subsection{Static model outputs}

Output CSV files for static models should contain columns with the following atmospheric quantities, in line with Table \ref{tab:static_params}:

\onecolumngrid
\begin{verbatim}
    z(m)            Height in atmosphere, starting from 0 (= surface)
    p_tot(bar)      Total atmospheric pressure at height z
    T(K)            Temperature at height z
    p_H2O(bar)      Partial atmospheric pressure of H2O at height z
    p_CO2(bar)      Partial atmospheric pressure of CO2 at height z
    p_CO(bar)       Partial atmospheric pressure of CO at height z
    p_H2(bar)       Partial atmospheric pressure of H2 at height z
    p_CH4(bar)      Partial atmospheric pressure of CH4 at height z
    p_O2(bar)       Partial atmospheric pressure of O2 at height z
\end{verbatim}

Some of these quantities are restating their input values, which will ensure simpler processing at later stages of the intercomparison. Some static models (e.g., LavaAtmos) only report surface quantities. These models should still follow the output above, but limit their output to one row. As before, quantities (columns) that are not computed by some of the codes should still be existent in the output file, but \rev{filled with NaN}.

Submitted output files for the static models should be:
\begin{itemize}
    \item \texttt{static-[modelname]-[planet]-tau[3-9]-[hot,cold]-data.csv}
    \begin{itemize}
        \item  For example: \texttt{static-moachi-trappist1b-tau5-hot-data.csv} for the output data of the static code MOAChi at age $\tau_5 = 10^5$ yr for the evolutionary model of TRAPPIST-1\,b with the hottest $T_\mathrm{surf}$ at this time.
    \end{itemize}
    \item \texttt{static-[modelname]-notes.txt}
    \item Any code configuration files necessary to recreate the output data.
\end{itemize}
\twocolumngrid

\bibliography{main}{}
\bibliographystyle{aasjournalv7}

\end{document}